\documentclass[a4paper,UKenglish,cleveref,autoref,thm-restate]{lipics-v2021}

\title{A Decidable Equivalence for a Turing-complete, Distributed Model of Computation} 

\titlerunning{A Decidable Equivalence} 

\author{Arnaldo Cesco}{Dipartimento di Informatica --- Scienza e Ingegneria, Universit\`a di Bologna, Mura A. Zamboni 7,40127 Bologna, Italy}{arnaldo.cesco@studio.unibo}{0000-0002-3417-1890}{}

\author{Roberto Gorrieri}{Dipartimento di Informatica --- Scienza e Ingegneria, Universit\`a di Bologna, Mura A. Zamboni 7,40127 Bologna, Italy}{roberto.gorrieri@unibo.it}{0000-0001-5502-0584}{}

\authorrunning{A. Cesco and R. Gorrieri} 

\Copyright{Arnaldo Cesco and Roberto Gorrieri} 



\relatedversiondetails[cite=pti-place-arxiv]{Previous Version}{https://arxiv.org/abs/2104.14859} 

\nolinenumbers 

\EventEditors{Filippo Bonchi and Simon J. Puglisi}
\EventNoEds{2}
\EventLongTitle{46th International Symposium on Mathematical Foundations of Computer Science (MFCS 2021)}
\EventShortTitle{MFCS 2021}
\EventAcronym{MFCS}
\EventYear{2021}
\EventDate{August 23--27, 2021}
\EventLocation{Tallinn, Estonia}
\EventLogo{}
\SeriesVolume{202}
\ArticleNo{28}

\usepackage{tikz}
\usetikzlibrary{arrows,decorations.pathmorphing,backgrounds,positioning,fit,petri}
\usepackage{centernot}

\newcommand{\post}[1]{\mbox{$#1^{\bullet}$}}
\newcommand{\pre}[1]{\mbox{$^{\bullet}#1$}}

\newcommand{\prei}[1]{\mbox{$^{\circ}#1$}}

\newcommand{\deriv}[1]{{\mbox{${\:\stackrel{#1}{\longrightarrow}\:}$}}}

\newcommand{\nat}{{\mathbb N}}
\newcommand{\dom}{\mathit{dom}}

\newcommand{\bigfrac}[2]{
\renewcommand{\arraystretch}{1.5}
\begin{array}{c}#1\\
\hline
#2
\end{array}}

\newcommand{\cntz}[0]{(\rho_1^0, C^0, \rho_2^0)}
\newcommand{\cnt}[0]{(\rho_{1}, C, \rho_{2})}
\newcommand{\cntp}[0]{(\rho_{1}', C', \rho_{2}')}

\newcommand{\trns}[1]{[{#1}\rangle}

\begin{document}

\maketitle

\begin{abstract}
Place/Transition Petri nets with inhibitor arcs (PTI nets for short), which are a well-known Turing-complete, distributed model of computation, 
are equipped with a decidable, behavioral equivalence, called pti-place bisimilarity, that conservatively extends place bisimilarity defined over 
Place/Transition nets (without inhibitor arcs). We prove that pti-place bisimilarity is sensible, as it respects the causal semantics of PTI nets.
\end{abstract}

\section{Introduction}

Place/Transition Petri nets with inhibitor arcs (PTI nets, for short), originally introduced in \cite{FA73}, are a well-known (see, e.g., \cite{tcsinib,Koutny,Pet81}), Turing-complete (as proved first by Agerwala in \cite{ager-pti}), distributed model of computation, largely exploited, e.g.,  
for modeling systems with priorities \cite{Hack}, for performance evaluation of distributed systems \cite{ajmone} and 
to provide $\pi$-calculus \cite{MPW,SW} with a 
net semantics \cite{BG09}.

As finite PTI nets constitute a Turing-complete model of computation, essentially all the properties of interest are undecidable, notably the reachability problem,
 and so even termination: it is undecidable whether a deadlock marking is reachable from the initial one.
 Also interleaving bisimulation equivalence is undecidable
 for finite PTI nets, as it is already undecidable \cite{Jan95} on the subclass of finite P/T nets \cite{Reisig}. Similarly, one can 
 prove that also well-known truly-concurrent behavioral equivalences, such as
 {\em fully-concurrent} bisimilarity \cite{BDKP91}, are undecidable \cite{Esp98} for finite PTI nets.
Despite this, we show that it is possible to define a {\em sensible}, behavioral equivalence which is actually {\em decidable} on finite PTI nets. 
This equivalence,
we call {\em pti-place bisimilarity}, is a conservative extension of {\em place bisimilarity on finite P/T nets}, introduced in \cite{ABS91} as an 
improvement of {\em strong bisimulation} \cite{Old}, 
(a relation proposed by Olderog in \cite{Old} on safe nets which fails to induce an equivalence relation), 
and recently proved decidable in \cite{Gor21}. 

Place bisimilarity on finite P/T nets is an equivalence over markings, based on relations over the {\em finite set of net places}, 
rather than over the (possibly infinite) set
of net markings. This equivalence is very natural and intuitive: 
as a place can be interpreted as a sequential process type (and each token
in this place as an instance of a sequential process of that type), a place bisimulation 
states which kinds of sequential processes (composing the distributed system represented by the finite P/T net)
are to be considered as equivalent. 
Moreover, this equivalence does respect the causal behavior of P/T nets, as van Glabbeek proved in \cite{G15} that 
it is slightly finer than 
{\em structure preserving bisimilarity} \cite{G15},
in turn slightly finer than {\em fully-concurrent bisimilarity} \cite{BDKP91}.

We extend this idea in order to be applicable to PTI nets. 
Informally, a binary relation $R$ over the set $S$ of places  is a {\em pti-place bisimulation} if for all markings $m_1$ and $m_2$
which are {\em bijectively} related via $R$ (denoted by $(m_1, m_2) \in R^\oplus$, 
where $R^\oplus$ is called the {\em additive closure} of $R$), 
if $m_1$ can perform transition $t_1$, reaching marking $m_1'$,
then $m_2$ can perform a transition $t_2$, reaching $m_2'$, such that 
\begin{itemize}
\item the pre-sets of $t_1$ and $t_2$ are related by $R^\oplus$,
the label of $t_1$ and $t_2$ is the same, the post-sets of $t_1$ and $t_2$ are related by $R^\oplus$, and also $(m_1', m_2') \in R^\oplus$,
as required by a place bisimulation \cite{ABS91,Gor21}, but additionally it is required that 
\item
whenever $(s, s') \in R$, 
$s$ belongs to the inhibiting set of $t_1$ if and only if $s'$  belongs to the inhibiting set of $t_2$;
\end{itemize}
and symmetrically if $m_2$ moves first. 
Two markings $m_1$ and $m_2$ are pti-place bisimilar, denoted by $m_1 \sim_p m_2$, if  a pti-place bisimulation $R$ exists such
that $(m_1, m_2) \in R^\oplus$.

We prove that pti-place bisimilarity is an equivalence, but it is not coinductive as the union of pti-place bisimulations 
may be not a pti-place bisimulation; so, in general, there is not a largest pti-place bisimulation, rather 
many maximal pti-place bisimulations.
In fact, pti-place bisimilarity is the relation on markings given by the union of the additive closure of each maximal pti-place bisimulation.
We also prove that $\sim_p$ is sensible, as it respects the causal semantics of PTI nets. 
As a matter of fact, 
following the approach in \cite{BP99,BP00}, we define a novel, process-oriented, bisimulation-based, behavioral semantics for PTI nets, called 
{\em causal-net bisimilarity}, and we prove that this is slightly coarser than pti-place bisimilarity.

The other main contribution of this paper is to show that $\sim_p$ is decidable for finite PTI nets.
As a place relation $R \subseteq S \times S$ is finite if the set $S$ of places is finite,
there are finitely many place relations for a finite net. We can list all these place relations, say $R_1, R_2, \ldots R_n$. 
It is possible to decide whether $R_i$ is a pti-place bisimulation by checking two {\em finite} conditions over 
a {\em finite} number of marking pairs: this is a non-obvious observation, as a pti-place bisimulation requires
that the pti-place bisimulation conditions hold for the infinitely many pairs $(m_1, m_2)$ belonging to $R_i^\oplus$. 
Hence, to decide whether 
$m_1 \sim_p m_2$, it is enough to check, for $i = $ $1, \ldots n$, whether $R_i$  is a pti-place 
bisimulation and, in such a case, whether $(m_1, m_2) \in R_i^\oplus$.

The paper is organized as follows. \cref{def-sec} recalls the basic definitions about PTI nets, including
their causal semantics.
\cref{iplace-sec} deals with pti-place bisimilarity, shows that it is an equivalence relation, that it is not coinductive,
and that it is slightly finer than causal-net bisimilarity.
\cref{decid-iplace-sec} shows that $\sim_p$ is decidable.
Finally, \cref{conc-sec} discusses some related literature 
and future research.

\section{Basic definitions about P/T nets and PTI nets}\label{def-sec}

 \begin{definition}\label{multiset}{\bf (Multiset)}\index{Multiset}
Let $\nat$ be the set of natural numbers. 
Given a finite set $S$, a {\em multiset} over $S$ is a function $m: S \rightarrow\nat$. 
The {\em support} set $\dom(m)$ of $m$ is $\{ s \in S \mid m(s) \neq 0\}$. 
The set of all multisets 
over $S$,  denoted by ${\mathcal M}(S)$, is ranged over by $m$. 
We write $s \in m$ if $m(s)>0$.
The {\em multiplicity} of $s$ in $m$ is given by the number $m(s)$. The {\em size} of $m$, denoted by $|m|$,
is the number $\sum_{s\in S} m(s)$, i.e., the total number of its elements.
A multiset $m$ such 
that $\dom(m) = \emptyset$ is called {\em empty} and is denoted by $\theta$.
We write $m \subseteq m'$ if $m(s) \leq m'(s)$ for all $s \in S$. 
{\em Multiset union} $\_ \oplus \_$ is defined as follows: $(m \oplus m')(s)$ $ = m(s) + m'(s)$. 
{\em Multiset difference} $\_ \ominus \_$ is defined as follows: 
$(m_1 \ominus m_2)(s) = max\{m_1(s) - m_2(s), 0\}$.
The {\em scalar product} of a number $j$ with $m$ is the multiset $j \cdot m$ defined as
$(j \cdot m)(s) = j \cdot (m(s))$. By $s_i$ we also denote the multiset with $s_i$ as its only element.
Hence, a multiset $m$ over $S = \{s_1, \ldots, s_n\}$
can be represented as $k_1\cdot s_{1} \oplus k_2 \cdot s_{2} \oplus \ldots \oplus k_n \cdot s_{n}$,
where $k_j = m(s_{j}) \geq 0$ for $j= 1, \ldots, n$.
\end{definition}

\begin{definition}\label{pt-net-def}{\bf (Place/Transition Petri net)}
A labeled, finite {\em Place/Transition} Petri net (P/T net for short) is a tuple $N = (S, A, T)$, where
\begin{itemize}
\item 
$S$ is the finite set of {\em places}, ranged over by $s$ (possibly indexed),
\item 
$A$ is the finite set of {\em labels}, ranged over by $\ell$ (possibly indexed), and
\item 
$T \subseteq ({\mathcal M}(S) \setminus \{\theta\}) \times A \times ({\mathcal M}(S) \setminus \{\theta\})$ 
is the finite set of {\em transitions}, 
ranged over by $t$ (possibly indexed).
\end{itemize}
Given a transition $t = (m, \ell, m')$,
we use the notation:
\begin{itemize}
\item  $\pre t$ to denote its {\em pre-set} $m$ (which cannot be empty) of tokens to be consumed;
\item $l(t)$ for its {\em label} $\ell$, and
\item $\post t$ to denote its {\em post-set} $m'$ (which cannot be an empty multiset) of tokens to be produced.
\end{itemize} 
Hence, transition $t$ can be also represented as $\pre t \deriv{l(t)} \post t$.
We also define the {\em flow function}
${\mbox  flow}: (S \times T) \cup (T \times S) \rightarrow \nat$ as follows:
for all $s \in S$, for all $t \in T$,
${\mbox  flow}(s,t) = \pre{t}(s)$ and ${\mbox  flow}(t,s) = \post{t}(s)$.
We will use $F$ to denote the {\em flow relation} 
$\{(x,y) \in (S \times T) \cup (T \times S) \mid {\mbox  flow}(x,y) > 0\}$.
Finally, we define pre-sets and post-sets also for places as follows: $\pre s = \{t \in T \mid s \in \post t\}$
and $\post s = \{t \in T \mid s \in \pre t\}$. 
Note that while the pre-set (post-set) of a transition is, in general, 
a multiset, the pre-set (post-set) of a place is a set.
\end{definition}

\begin{definition}\label{cpt-net-def}
{\bf (Place/Transition net with inhibitor arcs)}
A finite Place/Transition net {\em  with inhibitor arcs} (PTI net for short) is a tuple $N = (S, A, T, I)$, where
\begin{itemize}
    \item $(S, A, T)$ is a finite P/T net;
    \item $I \subseteq S \times T$ is the {\em inhibiting relation}.
    \end{itemize}
Given a transition $t \in T$, we denote by $\prei t$ its {\em inhibiting set} $\{ s \in S \mid (s, t) \in I \}$
 of places to be tested for absence of tokens. 
 Hence, a transition $t$ can be also represented as $(\pre t, \prei t) \deriv{l(t)} \post t$.
\end{definition}

We use the standard graphical convention for Petri nets.
In particular, a pair $(s, t)$ in the inhibiting relation $I$ is graphically represented by an arc from $s$ to $t$
ending with a small circle on the transition side.

\begin{definition}\label{pti-net-system}{\bf (Marking, PTI net system)}
A {\em PTI net system} $N(m_0)$ is a tuple $(S, A, T, I,$ $m_{0})$, 
where $(S,A, T, I)$ is a PTI net and $m_{0}$ is a multiset over $S$, called
the {\em initial marking}. We also say that $N(m_0)$ is a {\em marked} net.
\end{definition}

\begin{definition}\label{token-game} {\bf (Token game)}
A transition $t $ is {\em enabled} at $m$, denoted $m[t\rangle$, if $\pre t \subseteq m$
and $\prei t \cap \dom (m) = \emptyset$. 
The execution, or {\em firing}, of  $t$ enabled at $m$ produces the marking 
$m' = (m \ominus  \pre t) \oplus \post t$, written $m[t\rangle m'$. 
\end{definition}

\begin{definition}\label{net-system}{\bf (Firing sequence, reachable marking, safe net)}
A {\em firing sequence} starting at $m$ is defined inductively as follows:
\begin{itemize}
\item $m[\epsilon\rangle m$ is a firing sequence (where $\epsilon$ denotes an empty sequence of transitions) and
\item if $m[\sigma\rangle m'$ is a firing sequence and $m' [t\rangle m''$, then
$m [\sigma t\rangle m''$ is a firing sequence. 
\end{itemize}
The set of {\em reachable markings} from $m$ is 
$[m\rangle = \{m' \mid \exists \sigma.
m[\sigma\rangle m'\}$. 
A PTI system $N = $ $(S, A, T, I, m_0)$ is  {\em safe} if  for each marking 
$m \in [m_0\rangle$, we have that $m(s) \leq 1$ for all $s \in S$.
\end{definition}

\subsection{Causal semantics for P/T nets and PTI nets}
We outline some definitions about the causal semantics of P/T nets, adapted from the literature
(cf., e.g., \cite{BDKP91,G15,GR83,Old}).

\begin{definition}\label{acyc-def}{\bf (Acyclic net)}
A P/T net $N = (S, A, T)$ is
 {\em acyclic} if its flow relation $F$ is acyclic (i.e., $\nexists x$ such that $x \mathrel{F^+} x$,
 where $F^+$ is the transitive closure of $F$).
\end{definition}

The concurrent semantics of a marked P/T net is defined by a class of particular acyclic safe nets, 
where places are not branched (hence they represent a single run) and all arcs have weight 1. 
This kind of net is called {\em causal net}. 
We use the name $C$ (possibly indexed) to denote a causal net, the set $B$ to denote its 
places (called {\em conditions}), the set $E$ to denote its transitions 
(called {\em events}), and
$L$ to denote its labels.

\begin{definition}\label{causalnet-def}{\bf (Causal P/T net)}
A causal net is a finite marked net $C(\mathsf{m}_0) = (B,L, 
E,  \mathsf{m}_0)$ satisfying
the following conditions:
\begin{enumerate}
\item $C$ is acyclic;
\item $\forall b \in B \; \; | \pre{b} | \leq 1\, \wedge \, | \post{b} | \leq 1$ (i.e., the places are not branched);
\item  $ \forall b \in B \; \; \mathsf{m}_0(b)   =  \begin{cases}
 1 & \mbox{if $\; \pre{b} = \emptyset$}\\ 
  0  & \mbox{otherwise;}   
   \end{cases}$\\
\item $\forall e \in E \; \; \pre{e}(b) \leq 1 \, \wedge \, \post{e}(b) \leq 1$ for all $b \in B$ (i.e., all the arcs have weight $1$).
\end{enumerate}
We denote by $Min(C)$ the set $\mathsf{m}_0$, and by $Max(C)$ the set
$\{b \in B \mid \post{b} = \emptyset\}$.
\end{definition}

Note that any reachable marking of a causal net is a set, i.e., 
this net is {\em safe}; in fact, the initial marking is a set and, 
assuming by induction that a reachable marking $\mathsf{m}$ is a set and enables $e$, i.e., 
$\mathsf{m}[e\rangle \mathsf{m}'$,
then also
$\mathsf{m}' =  (\mathsf{m} \ominus \pre{e}) \oplus \post{e}$ is a set, 
as the net is acyclic and because
of the condition on the shape of the post-set of $e$ (weights can only be $1$).

As the initial marking of a causal P/T net is fixed by its shape (according to item $3$ of 
\cref{causalnet-def}), in the following, in order to make the 
notation lighter, we often omit the indication of the initial marking (also in their graphical representation), so that the causal 
net $C(\mathsf{m}_0)$ is denoted by $C$.

\begin{definition}\label{trans-causal}{\bf (Moves of a causal P/T net)}
Given two causal nets $C = (B, L, E,  \mathsf{m}_0)$
and $C' = (B', L, E',  \mathsf{m}_0)$, we say that $C$
moves in one step to $C'$ through $e$, denoted by
$C [e\rangle C'$, if $\; \pre{e} \subseteq Max(C)$, $E' = E \cup \{e\}$
and $B' = B \cup \post{e}$. 
\end{definition}

\begin{definition}\label{folding-def}{\bf (Folding and Process)}
A {\em folding} from a causal P/T net $C = (B, L, E, \mathsf{m}_0)$ into a P/T net system
$N(m_0) = (S, A, T, m_0)$ is a function $\rho: B \cup E \to S \cup T$, which is type-preserving, i.e., 
such that $\rho(B) \subseteq S$ and $\rho(E) \subseteq T$, satisfying the following:
\begin{itemize}
\item $L = A$ and $\mathsf{l}(e) = l(\rho(e))$ for all $e \in E$;
\item $\rho(\mathsf{m}_0) = m_0$, i.e., $m_0(s) = | \rho^{-1}(s) \cap \mathsf{m}_0 |$;
\item $\forall e \in E, \rho(\pre{e}) = \pre{\rho(e)}$, i.e., $\rho(\pre{e})(s) = | \rho^{-1}(s) \cap \pre{e} |$
for all $s \in S$;
\item $\forall e \in E, \, \rho(\post{e}) = \post{\rho(e)}$,  i.e., $\rho(\post{e})(s) = | \rho^{-1}(s) \cap \post{e} |$
for all $s \in S$.
\end{itemize}
A pair $(C, \rho)$, where $C$ is a causal net and $\rho$ a folding from  
$C$ to a net system $N(m_0)$, is a {\em process} of $N(m_0)$.
\end{definition}

\begin{definition}\label{trans-process}{\bf (Moves of a P/T process)}
Let $N(m_0) = (S, A, T, m_0)$ be a net system 
and let $(C_i, \rho_i)$, for $i = 1, 2$, be two processes of $N(m_0)$.
We say that $(C_1, \rho_1)$
moves in one step to $(C_2, \rho_2)$ through $e$, denoted by
$(C_1, \rho_1) \deriv{e} (C_2, \rho_2)$, if $C_1 [e\rangle C_2$
and $\rho_1 \subseteq \rho_2$.
\end{definition}

Following \cite{BP99,BP00}, we define here a possible causal semantics for PTI nets. 
In order to maintain the pleasant property that a process univocally determines
the causal dependencies among its events, it is not enough to just enrich 
causal P/T nets with inhibitor arcs. 
Indeed, the reason why a condition is empty may influence the causal relation
of events. 
To solve the problem, in \cite{BP99,BP00} inhibitor arcs are partitioned into two 
sets: {\em before} inhibitor arcs and {\em after} inhibitor arcs. 
If a condition is connected to an event by a before inhibitor arc,
the event fires because the condition has not held yet; 
if they are connected by an after inhibitor arc, 
the event fires because the condition does not hold anymore.

\begin{definition}\label{causal-pti-net-def}{\bf (Causal PTI net)}
A causal PTI net is a tuple
$C(\mathsf{m}_0) = (B,L, E, Y^{be}, Y^{af},$ $\mathsf{m}_0)$
satisfying the following conditions, denoting the flow relation of $C$ by $\mathsf{F}$:
\begin{enumerate}
\item $(B,L,E, \mathsf{m}_0)$ is a causal P/T net;
\item $(B,L, E, Y^{be} \cup Y^{af}, \mathsf{m}_0)$ is a marked PTI net;
\item {\em before} and {\em after} requirements are met, i.e.
    \begin{alphaenumerate}
        \item If $b \mathrel{Y^{be}} e$, then there exists $e' \in E$ such that $e' \mathrel{\mathsf{F}} b$, and
        \item If $b \mathrel{Y^{af}} e$, then there exists $e' \in E$ such that $b \mathrel{\mathsf{F}} e'$;
    \end{alphaenumerate}
\item relation $\mathsf{F}\, \cup \prec_{af} \cup \prec_{be}$ is acyclic, where
$\prec_{af} = \mathsf{F}^{-1} \circ Y^{af}$ and
$\prec_{be} = (Y^{be})^{ -1} \circ \mathsf{F}^{-1}$.
\end{enumerate}
We denote by $Min(C)$ the set $\mathsf{m}_0$, and by $Max(C)$ the set
$\{b \in B \mid \post{b} = \emptyset\}$.
\end{definition}

Relation $\prec_{af} \subseteq E \times E$ states that $e \prec_{af} e'$ if $e$ consumes the token in a 
place $b$ inhibiting $e'$: this is clearly a causal dependency.
Instead, relation $\prec_{be} \subseteq E \times E$ states that $e \prec_{be} e'$ if $e'$ produces 
a token in a place $b$ inhibiting $e$: this is clearly a 
temporal precedence, because the two events can be causally independent, yet they cannot 
occur in any order, as if $e'$ occurs, then $e$ is disabled.
 
 \begin{definition}\label{folding-pti--def}{\bf (Folding and PTI process)}
A {\em folding} from a causal PTI net $C = (B, L, E,$ $Y^{be}, Y^{af}, \mathsf{m}_0)$
into a PTI net system
$N(m_0) = (S, A, T, I, m_0)$
is a function $\rho: B \cup E \to S \cup T$,
which is type-preserving, i.e.,
such that $\rho(B) \subseteq S$ and $\rho(E) \subseteq T$,
satisfying the following:
\begin{itemize}
\item $\rho$ is a P/T folding from $(B,L,E, \mathsf{m}_0)$ into $(S, A, T, m_0)$;

\item for all $s \in S$ and $e \in E$, if $(s,\rho(e)) \in I$ then for all
        $b \in B$ such that $\rho(b) = s$, it holds
        $(b, e) \in Y^{be} \cup Y^{af} \cup \mathsf{F}^{-1}$, and \\
        for all $b \in B$ and $e \in E$, if $(b, e) \in Y^{be} \cup Y^{af}$
        then $(\rho(b), \rho(e)) \in I$.
\end{itemize}
A pair $(C, \rho)$, where $C$ is a causal PTI net and $\rho$ a folding from  
$C$ to a PTI net system $N(m_0)$, is a {\em PTI process} of $N(m_0)$.
\end{definition}

Each inhibitor arc in the causal net has a corresponding inhibitor arc 
in the net system.
The only case wherea condition $b$ is not connected by an inhibitor arc to an event $e$ 
is when $b$ is in the post-set of $e$: as $b$ starts to hold only after $e$
occurs, the only possibility is to put a before arc. This would make the 
relation $\prec_{be}$ reflexive, invalidating item 4 of \cref{causal-pti-net-def}.
However, since $b$ is in the post-set of $e$, we are sure that $e$ happens before $b$
is fulfilled, hence making useless the presence of a before inhibitor arc. 
For this reason, with the requirement $(b, e) \in Y^{be} \cup Y^{af} \cup \mathsf{F}^{-1}$, 
we ask for the presence of an inhibitor arc only if there exists no flow from
$e$ to $b$.

\begin{definition}\label{trans-pti-process}{\bf (Moves of a PTI process)}
Let $N(m_0) = (S, A, T, I, m_0)$ be a PTI net system 
and let $(C_i, \rho_i)$, for $i = 1, 2$, be two PTI processes of $N(m_0)$, 
where $C_i = (B_i, L, E_i,$  $Y_i^{be}, Y_i^{af}, \mathsf{m}_0) $.
We say that $(C_1, \rho_1)$
moves in one step to $(C_2, \rho_2)$ through $e$, denoted by
$(C_1, \rho_1) \deriv{e} (C_2, \rho_2)$, 
if the following hold:

\begin{itemize}
    \item $\pre{e} \subseteq Max(C_1)$, $E_2 = E_1 \cup \{ e \}$, $B_2 = B_1 \cup \post{e}$, $\rho_1 \subseteq \rho_2$,
    i.e. the P/T process of $(C_1, \rho_1)$ moves in one step through $e$ to the P/T process of $(C_2, \rho_2)$.
        
    \item Given two relations $\mathcal{B}$ and $\mathcal{A}$, defined as
        \begin{itemize}
            \item[-] $\forall b \in \post{e}$, $\forall e' \in E_1$ 
            we have $\;b \mathrel{\mathcal{B}} e'\;$ if and only if $(\rho_2(b), \rho_2(e')) \in I$,
            
            \item[-] $\forall b \in B_2$ such that $\post{b} \neq \emptyset$, we have $\; b \mathrel{\mathcal{A}} e\;$ if and only if $(\rho_2(b), \rho_2(e)) \in I$,
        \end{itemize}
        we have
        $\{ b \in B_2 \mid b \mathrel{\mathcal{A}} e\} \cap 
        Max(C_1) = \emptyset$.
    
    \item Finally, $Y_2^{be} = Y_1^{be} \cup \mathcal{B}$ and $Y_2^{af} = Y_1^{af} \cup \mathcal{A}$.
\end{itemize}
\end{definition}

The item 
$ \{ b \in B_2 \mid b \mathrel{\mathcal{A}} e\} \cap 
        Max(C_1) = \emptyset$
models the fact that a transition can fire only
if all its inhibiting places are free. Indeed,
an event can fire only if its (so far known) 
inhibiting conditions are not maximal.
Note that, by construction, before arcs can connect
only new inhibiting conditions 
to past events and in particular we
do not allow before arcs
connecting a condition in the post-set of a newly added event $e$
with the event $e$ itself. 
Moreover, after arcs can only connect old inhibiting conditions
to the new event $e$ and since 
$ \{ b \in B_2 \mid b \mathrel{\mathcal{A}} e\} \cap Max(C_1) = \emptyset$,
the old inhibiting conditions cannot be in the pre-set of the newly added event $e$.
Therefore, both relations $\prec_2^{be}$ and $\prec_2^{af}$ are acyclic, 
and since $\mathsf{F}_2$ is acyclic too, $(C_2, \rho_2)$ is truly a process of $N(m_0)$.

\begin{figure}
    \centering
    \begin{tikzpicture}[
        every place/.style={draw,thick,inner sep=0pt,minimum size=6mm},
        every transition/.style={draw,thick,inner sep=0pt,minimum size=4mm},
        bend angle=30,
        pre/.style={<-,shorten <=1pt,>=stealth,semithick},
        post/.style={->,shorten >=1pt,>=stealth,semithick}
    ]
    \def\eofigdist{3.3cm}
    \def\eodist{0.35cm}
    \def\eodisty{0.5cm}

\node (N) [label=$N)$]{};
    
\node (s1) [place, tokens=1]  [right=\eodist of N, label=above:$s_1$] {};
\node (t1) [transition] [below=\eodisty of s1, label=right:$t_1$] {};
\node (s2) [place]  [below=\eodisty of t1, label=below:$s_2$] {};
\node (s3) [place, tokens=1]  [right=\eodisty of s1, label=above:$s_3$] {};
\node (t2) [transition] [below=\eodisty of s3, label=right:$t_2$] {};
\node (s4) [place]  [below=\eodisty of t2, label=below:$s_4$] {};
\node (t3) [transition] [left=\eodist of s2, label=below:$t_3$] {};
\node (s5) [place]  [left=\eodist of t3, label=below:$s_5$] {};

\draw  [->] (s1) to (t1);
\draw  [->] (t1) to (s2);
\draw  [-o] (s2) to (t2);
\draw  [->] (s3) to (t2);
\draw  [->] (t2) to (s4);
\draw  [->] (s2) to (t3);
\draw  [->] (t3) to (s5);

\node (C1) [right={4.2cm} of N, label=$C_1)$]{};

\node (b1) [place]  [right=\eodist of C1, label=above:$b_1$] {};
\node (e1) [transition] [below=\eodisty of b1, label=right:$e_1$] {};
\node (b2) [place]  [below=\eodisty of e1, label=below:$b_2$] {};
\node (b3) [place]  [right=\eodisty of b1, label=above:$b_3$] {};
\node (e2) [transition] [below=\eodisty of b3, label=right:$e_2$] {};
\node (b4) [place]  [below=\eodisty of e2, label=below:$b_4$] {};
\node (e3) [transition] [left=\eodist of b2, label=below:$e_3$] {};
\node (b5) [place]  [left=\eodist of e3, label=below:$b_5$] {};

\draw  [->] (b1) to (e1);
\draw  [->] (e1) to (b2);
\draw  (b2) -o (e2) node[midway,right, rotate=0] {$a$};
\draw  [->] (b3) to (e2);
\draw  [->] (e2) to (b4);
\draw  [->] (b2) to (e3);
\draw  [->] (e3) to (b5);

\node (C2) [right={4.2cm} of C1, label=$C_2)$]{};

\node (b1) [place]  [right=\eodist of C2, label=above:$b_1$] {};
\node (e1) [transition] [below=\eodisty of b1, label=right:$e_1$] {};
\node (b2) [place]  [below=\eodisty of e1, label=below:$b_2$] {};
\node (b3) [place]  [right=\eodisty of b1, label=above:$b_3$] {};
\node (e2) [transition] [below=\eodisty of b3, label=right:$e_2$] {};
\node (b4) [place]  [below=\eodisty of e2, label=below:$b_4$] {};
\node (e3) [transition] [left=\eodist of b2, label=below:$e_3$] {};
\node (b5) [place]  [left=\eodist of e3, label=below:$b_5$] {};

\draw  [->] (b1) to (e1);
\draw  [->] (e1) to (b2);
\draw  (b2) -o (e2) node[midway,right, rotate=0] {$b$};
\draw  [->] (b3) to (e2);
\draw  [->] (e2) to (b4);
\draw  [->] (b2) to (e3);
\draw  [->] (e3) to (b5);

\end{tikzpicture}
    
    \caption{A marked PTI net and two PTI causal nets corresponding to its two maximal processes.}
    \label{fig:1}
\end{figure}
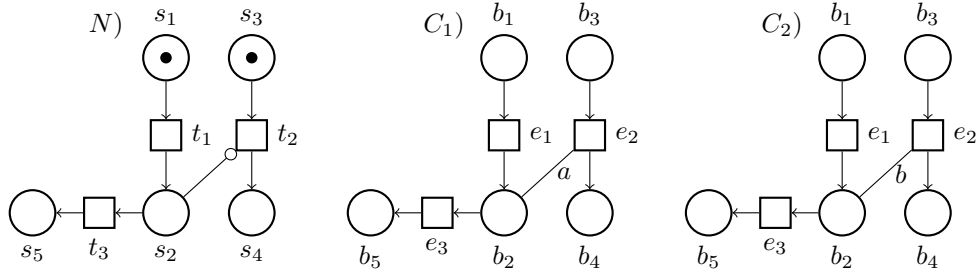

\begin{example}
Consider the three nets in \cref{fig:1}, where we use the graphical convention that
{\em before} inhibitor arcs and {\em after} inhibitor arcs are represented
by lines between a condition and an event: the former labeled by $b$, 
the latter labeled by $a$.
The initial marking of $N$ is $m_0 = s_1 \oplus s_3$.
The shape of a process generated by $N(m_0)$ may depend on the order of transitions in a
given transition sequence. 
As a matter of fact, transition sequences containing the same transitions but in 
a different order may generate different processes, e.g. $C_1$ and $C_2$.
Indeed, $C_1$ represents the transition sequence $t_1 \, t_3 \, t_2$, while
$C_2$ represents the transition sequence $t_2 \, t_1 \, t_3$.
In the former case, $t_1$ fires first, so that $t_2$ can only fire after the 
inhibiting token in $s_2$ has been cleaned up by transition $t_3$: therefore
the causal net has an after arc between $b_2$ and $e_2$. 
In the latter case, $t_2$ is the first transition to fire and there are no tokens in
the inhibiting place $b_2$, therefore the causal net has a before arc between $b_2$ and $e_2$. 
Note that the underlying causal P/T net of these two processes is the same,
but before and after inhibitor arcs are different.
\end{example}

We are now ready to introduce a novel behavioral relation for PTI nets, namely causal-net bisimulation,
which is an interesting relation in its own right, as the induced equivalence, namely {\em causal-net bisimilarity}, 
on P/T nets coincides with {\em structure-preserving bisimilarity} \cite{G15},
and so it is slightly finer than {\em fully-concurrent bisimilarity} \cite{BDKP91}. However,
since we conjecture that causal-net bisimilarity is undecidable (already on finite P/T nets),
we will use this behavioral relation only for comparison with pti-place bisimilarity, showing the the latter is a finer, but decidable,
approximation of the former.

\begin{definition}\label{cn-bis-def}{\bf (Causal-net bisimulation)}
Let $N = (S, A, T, I)$ be a PTI net. 
A {\em causal-net bisimulation} 
is a relation $R$, composed of 
triples of the form $(\rho_1, C, \rho_2)$, where, for $i = 1, 2$, $(C, \rho_i)$ is a process of $N(m_{0i})$ for some $m_{0i}$,
such that if $(\rho_1, C, \rho_2) \in R$ then

\begin{itemize}
\item[$i)$] 
$\forall t_1, C', \rho_1'$ such that $(C, \rho_1) \deriv{e} (C', \rho_1')$,
where $\rho_1'(e) = t_1$,
$\exists t_2, \rho_2'$ such that\\
$(C, \rho_2) \deriv{e} (C', \rho_2')$,
where $\rho_2'(e) = t_2$, and
$(\rho'_1, C', \rho'_2) \in R$;

\item[$ii)$] symmetrically, 
$\forall t_2, C', \rho_2'$ such that $(C, \rho_2) \deriv{e} (C', \rho_2')$,
where $\rho_2'(e) = t_2$,
$\exists t_1, \rho_1'$ such that
$(C, \rho_1) \deriv{e} (C', \rho_1')$,
where $\rho_1'(e) = t_1$, and
$(\rho'_1, C', \rho'_2) \in R$.

\end{itemize}

Two markings $m_{1}$ and $m_2$ of $N$ are cn-bisimilar, 
denoted by $m_{1} \sim_{cn} m_{2}$, 
if there exists a causal-net bisimulation $R$ containing a triple $(\rho^0_1, C^0, \rho^0_2)$, 
where $C^0$ contains no events and 
$\rho^0_i(Min( C^0))  = \rho^0_i(Max( C^0)) = m_i\;$ for $i = 1, 2$.
\end{definition}

If $m_1 \sim_{cn} m_2$, then these two markings have the same causal PTI nets,
so that the executions originating from the two markings have the same 
causal dependencies (determined by $\mathsf{F}$ and $\prec_{af}$) and 
the same temporal dependencies (determined by $\prec_{be}$).
Causal-net bisimilarity $\sim_{cn}$ is an equivalence relation 
(see the preliminary version of this paper \cite{pti-place-arxiv}).

\section{Pti-place bisimilarity}\label{iplace-sec}
We now present pti-place bisimilarity, 
which conservatively extends
{\em place bisimilarity} \cite{ABS91,Gor21} to the case of PTI nets.
First, an auxiliary definition.

\subsection{Additive closure and its properties}

\begin{definition}\label{add-eq}{\bf (Additive closure)}
Given a PTI net $N = (S, A, T, I)$ and a {\em place relation} $R \subseteq S \times S$, we define a {\em marking relation}
$R^\oplus \, \subseteq \, {\mathcal M}(S) \times {\mathcal M}(S)$, called 
the {\em additive closure} of $R$,
as the least relation induced by the following axiom and rule.\\

$\begin{array}{lllllllllll}
 \bigfrac{}{(\theta, \theta) \in  R^\oplus} & \;   &   \; 
 \bigfrac{(s_1, s_2) \in R \;  (m_1, m_2) \in R^\oplus }{(s_1 \oplus m_1, s_2 \oplus m_2) \in  R^\oplus }  \\
\end{array}$
\end{definition}

Note that  two markings are related by $R^\oplus$ only if they have the same size; 
in fact, the axiom states that
the empty marking is related to itself, while the rule, assuming by induction 
that $m_1$ and $m_2$ have the same size, ensures that $s_1 \oplus m_1$ and
$s_2 \oplus m_2$ have the same size.

\begin{proposition}\label{fin-k-add}
For each relation $R \subseteq S \times S$,  if $(m_1, m_2) \in R^\oplus$, 
then $|m_1| = |m_2|$.
\end{proposition}

Note also that the membership $(m_1, m_2) \in R^\oplus$ may be proved in several different ways, 
depending on the chosen order of the elements
of the two markings and on the definition of $R$. For instance, if $R = \{(s_1, s_3),$ $(s_1, s_4),$ $(s_2, s_3), (s_2, s_4)\}$,
then $(s_1 \oplus s_2, s_3 \oplus s_4) \in R^\oplus$ can be proved by means of the pairs $(s_1, s_3)$ and $(s_2, s_4)$,
as well as by means of $(s_1, s_4), (s_2, s_3)$.
An alternative way to define that two markings $m_1$ and $m_2$
are related by $R^\oplus$ is to state that $m_1$ can be represented as $s_1 \oplus s_2 \oplus \ldots \oplus s_k$, 
$m_2$ can be represented as $s_1' \oplus s_2' \oplus \ldots \oplus s_k'$ and $(s_i, s_i') \in R$ for $i = 1, \ldots, k$. 
\begin{proposition}\label{add-prop1}\cite{Gor17b}
For each place relation $R \subseteq S \times S$, the following hold:
\begin{enumerate}
\item If $R$ is an equivalence relation, then $R^\oplus$ is an equivalence relation.
\item If $R_1 \subseteq R_2$, then $R_1^\oplus \subseteq R_2^\oplus$, i.e., the additive closure is monotone.
\item If $(m_1, m_2) \in R^\oplus$ and $(m_1', m_2') \in R^\oplus$,
then $(m_1 \oplus m_1', m_2 \oplus m_2') \in R^\oplus$, i.e., the additive closure is additive.
\item If $R$ is an equivalence relation and, moreover, $(m_1 \oplus m_1', m_2 \oplus m_2') \in  R^\oplus$ 
and $(m_1, m_2) \in R^\oplus$,
then $(m_1', m_2') \in R^\oplus$, i.e., the additive closure of an equivalence place relation is subtractive.
\end{enumerate}
\end{proposition}

When $R$ is an equivalence relation, it is rather easy to check whether two markings are related by $R^\oplus$. An 
algorithm, described in \cite{Gor17b}, establishes whether an $R$-preserving bijection exists between 
the two markings $m_1$ and $m_2$ of equal size $k$ in $O(k^2)$ time. 
Another algorithm,
described in \cite{Lib19,Gor20c}, 
checks whether $(m_1, m_2) \in R^\oplus$ in $O(n)$ time, where $n$ is the size of $S$.
However, these performant algorithms heavily rely on the fact that $R$ is an equivalence relation, hence also 
subtractive (case 4 of Proposition \ref{add-prop1}).
If $R$ is not an equivalence relation, which is typical  for place bisimulations, the naive algorithm for checking 
whether $(m_1, m_2) \in R^\oplus$ would simply consider 
$m_1$ represented as $s_1 \oplus s_2 \oplus \ldots \oplus s_k$, and then would scan all the possible permutations of 
$m_2$, each represented as $s'_1 \oplus s'_2 \oplus \ldots \oplus s'_k$, 
to check that $(s_i, s_i') \in R$ for $i = 1, \ldots, k$. Of course, this naive algorithm has worst-case complexity $O(k!)$.

\begin{example}\label{nsubtractive}
Consider $R = \{(s_1, s_3),$  $(s_1, s_4), (s_2, s_4)\}$, which is not an equivalence relation.
Suppose we want to check that $(s_1 \oplus s_2, s_4 \oplus s_3) \in R^\oplus$.
If we start by matching $(s_1, s_4) \in R$, then we fail because the residual $(s_2, s_3)$ is not in $R$.
However, if we permute the second marking to $s_3 \oplus s_4$, then we succeed because the required pairs
$(s_1, s_3)$ and $(s_2, s_4)$ are both in $R$.
\end{example}

Nonetheless, the problem of checking whether $(m_1, m_2) \in R^\oplus$ has polynomial 
time complexity because it can be considered as an instance of
the problem of finding a perfect matching in a bipartite graph, 
where the nodes of the two partitions are the tokens in the 
two markings, and the edges
are defined by the relation $R$. 
In fact, 
the definition of the bipartite graph takes $O(k^2)$ time (where $k = |m_1| = |m_2|$) and, then, 
the Hopcroft-Karp-Karzanov algorithm  \cite{HK73} for computing the maximum matching has
worst-case time complexity $O(h\sqrt{k})$, where $h$ is the number of the edges in the bipartire graph ($h \leq k^2$) and
to check whether the maximum matching is perfect can be done simply 
by checking that the size of the matching equals the number of nodes in each partition, i.e., $k$.
Hence, in evaluating the complexity of the algorithm in Section \ref{decid-iplace-sec}, we assume that the complexity of 
checking whether $(m_1, m_2) \in R^\oplus$ is in $O(k^2 \sqrt{k})$.

A related problem is that of computing, given a marking $m_1$ of size $k$, the set of all the markings $m_2$ such that 
$(m_1, m_2) \in R^\oplus$. This problem can be solved with a worst-case time complexity of $O(n^k)$ because each of the $k$
tokens in $m_1$ can be related via $R$ to $n$ places at most.

Now we list some necessary, and less obvious, properties of additively closed place relations that will be useful in the following.

\begin{proposition}\label{add-prop2}\cite{Gor17b}
For each family of place relations $R_i \subseteq S \times S$, the following hold:
\begin{enumerate}
\item $\emptyset^\oplus = \{(\theta, \theta)\}$, i.e., the additive closure of the empty place relation
yields a singleton marking relation, relating the empty marking to itself.
\item $(\mathcal{I}_S)^\oplus = \mathcal{I}_M$, i.e., the additive closure of the
identity on places $\mathcal{I}_S = \{(s, s) \mid s \in S\}$ yields the identity relation on markings
$\mathcal{I}_M = \{(m, m) \mid $ $ m \in  {\mathcal M}(S)\}$.
\item $(R^\oplus)^{-1} = (R^{-1})^\oplus$, i.e., the inverse of an additively closed relation $R$ equals the additive closure
of its inverse $R^{-1}$.
\item $(R_1 \circ R_2)^\oplus = (R_1^\oplus) \circ (R_2^\oplus)$, i.e., the additive closure of the composition of two 
place relations equals the compositions of their additive closures.
\end{enumerate}
\end{proposition}

\subsection{Pti-place bisimulation and its properties}

We are now ready to introduce pti-place bisimulation, 
which is a non-interleaving behavioral relation 
defined over the net places.
Note that for P/T nets, place bisimulation \cite{ABS91,Gor21}
and pti-place bisimulation coincide because $I = \emptyset$.

\begin{definition}\label{def-pti-place-bis}{\bf (Pti-place bisimulation)}
Let $N = (S, A, T, I)$ be a PTI net. 
A {\em pti-place bisimulation} is a relation
$R\subseteq S \times S$ such that if $(m_1, m_2) \in R^\oplus$
then
\begin{enumerate}
\item $\forall t_1$ such that  $m_1[t_1\rangle m'_1$, $\exists t_2$ such that $m_2[t_2\rangle m'_2$ 
and
          \begin{alphaenumerate}
          \item $(\pre{t_1}, \pre{t_2}) \in R^\oplus$, , 
          $(\post{t_1}, \post{t_2}) \in R^\oplus$, $l(t_1) = l(t_2)$, and $(m_1 \ominus \pre{t_1}, m_1 \ominus \pre{t_2}) \in R^\oplus$,
          \item $\forall s, s' \in S. (s, s') \in R \Rightarrow (s \in \prei{t_1} \Leftrightarrow s' \in \prei{t_2})$.
          \end{alphaenumerate}

\item $\forall t_2$ such that  $m_2[t_2\rangle m'_2$, $\exists t_1$ such that $m_1[t_1\rangle m'_1$ 
and
          \begin{alphaenumerate}
          \item $(\pre{t_1}, \pre{t_2}) \in R^\oplus$, , 
          $(\post{t_1}, \post{t_2}) \in R^\oplus$, $l(t_1) = l(t_2)$, and $(m_1 \ominus \pre{t_1}, m_1 \ominus \pre{t_2}) \in R^\oplus$,
          \item $\forall s, s' \in S. (s, s') \in R \Rightarrow (s \in \prei{t_1} \Leftrightarrow s' \in \prei{t_2})$.
          \end{alphaenumerate}
\end{enumerate}

Two markings $m_1$ and $m_2$ are  {\em pti-place bisimilar}, denoted by
$m_1 \sim_p m_2$, if there exists a pti-place bisimulation $R$ such that $(m_1, m_2) \in R^\oplus$.
\end{definition}
Note that, by additivity of $R^\oplus$ (cf. Proposition \ref{add-prop1}), 
from $(m_1 \ominus \pre{t_1}, m_2 \ominus \pre{t_2}) \in R^\oplus$
and $(\post{t_1}, \post{t_2}) \in R^\oplus$ we derive $(m'_1, m'_2) \in R^\oplus$, 
which is  the condition required in the original definition of place
bisimulation in \cite{ABS91}. 
Our slightly stronger formulation is more adequate for the proof of \cref{pti-bis-finer-cn-bis}.

Conditions 1(b) and 2(b) make sure that the relation $R$ 
respects the inhibiting behavior of places. Indeed, 
an inhibiting place for one of the two transitions
cannot be related via $R$ to a non-inhibiting place for the 
other transition.
These conditions might appear rather restrictive, and one may wonder whether they can be weakened or omitted altogether. However, their presence is strictly necessary in the crucial step of the proof of \cref{r-pti-place-bis-decid}. Moreover, these conditions are also essential for proving \cref{pti-bis-finer-cn-bis}.

\begin{figure}[t]
    \centering
    \begin{tikzpicture}[
        every place/.style={draw,thick,inner sep=0pt,minimum size=6mm},
        every transition/.style={draw,thick,inner sep=0pt,minimum size=4mm},
        bend angle=30,
        pre/.style={<-,shorten <=1pt,>=stealth,semithick},
        post/.style={->,shorten >=1pt,>=stealth,semithick}
    ]
    \def\eofigdist{3.3cm}
    \def\eodist{0.5cm}
    \def\eodisty{0.75cm}

\node (N2) [label=$N_1)$]{};
{
    \node (s2) [place] [right={1.6cm} of N2, label=above:$s_2$]{};
    \node (t1) [transition] [below=\eodisty of s2, label=right:$a$] {};
    \node (s0) [place] [left=\eodisty of t1, label=left:$s_0$] {};
    \node (s3) [place] [below=\eodisty of t1, label=right:$s_3$] {};
    \node (t2) [transition] [below=\eodisty of s3, label=right:$b$] {};
    \node (s1) [place] [left=\eodisty of t2, label=left:$s_1$] {};

    \draw  [-o] (s0) to (t1);
    \draw  [->] (s2) to (t1);
    \draw  [->] (t1) to (s3);
    \draw  [-o] (s1) to (t2);
    \draw  [->] [bend right] (s3) to (t2);
    \draw  [->] [bend right] (t2) to node[auto,swap] {2} (s3);
    
    \node (s4) [place] [right={9cm} of N2, label=above:$s_4$]{};
    \node (t1) [transition] [below of = s4, label=right:$a$]{};
    \node (s5) [place] [left=\eodisty of t1, label=above:$s_5$]{};
    \node (t2) [transition] [left=\eodisty of s5, label=left:$a$]{};
    \node (s6) [place] [below=\eodisty of t1, label=below:$\qquad s_6$]{};
    \node (t3) [transition] [left=\eodisty of s6, label=above:$\quad a$]{};
    \node (s7) [place] [left=\eodisty of t3, label=below:$s_7$]{};
    \node (t5) [transition] [below=\eodisty of s6, label=below:$b$]{};
    \node (s10) [place] [left=\eodisty of t5, label=below:$s_{10}$]{};
    \node (t6) [transition] [left=\eodisty of s10, label=left:$b$]{};
    \node (t7) [transition] [left={1.5cm} of t6, label=left:$b$]{};
    \node (s8) [place] [above=\eodisty of t7, label=above:$s_8$]{};
    \node (s9) [place] [right=\eodisty of t5, label=above:$s_9$]{};
    \node (t4) [transition] [right=\eodisty of s9, label=right:$b$]{};
    \node (s11) [place] [below=\eodisty of s10, label=below:$s_{11}$]{};

    \draw  [-o] (s5) to (t1);
    \draw  [-o] (s5) to (t2);
    \draw  [-o] (s5) to (t3);
    \draw  [-o] [bend right] (s11) to (t4);
    \draw  [-o] (s11) to (t5);
    \draw  [-o] (s11) to (t6);
    \draw  [-o] [bend left] (s11) to (t7);
    \draw  [->] (s4) to (t1);
    \draw  [->] (t1) to (s6);
    \draw  [->] (s7) to (t2);
    \draw  [->] (s7) to (t3);
    \draw  [->] (t3) to (s6);
    \draw  [->] (t2) to (s8);
    \draw  [->] (s6) to (t5);
    \draw  [->] (t5) to (s10);
    \draw  [->] (t5) to (s9);
    \draw  [->] (s10) to (t6);
    \draw  [->] (s9) to (t4);
    \draw  [->] [bend right] (t4) to (s9);
    \draw  [->] [bend right] (t4) to (s6);
    \draw  [->] (t6) --node[fill=white,sloped] {2} (s6);                   
    \draw  [->] (s8) to (t7);
    \draw  [->] [bend right] (t7) to node[auto,swap] {2} (s8);
}

\node (N3) [below={6.5cm} of N2, label=$N_2)$]{};
{
    \node (s1) [place] [right=\eodist of N3, label=above:$s_1$] {};
    \node (t1) [transition] [below=\eodisty of s1, label=left:$a$] {};
    \node (s4) [place]  [below=\eodisty of t1, label=below:$s_4$] {};
    
    \node (s2) [place]  [right=\eodisty of s1, label=above:$s_2$] {};
    \node (t2) [transition] [below=\eodisty of s2, label=right:$b$] {};
    \node (s5) [place]  [below=\eodisty of t2, label=below:$s_5$] {};
    
    \node (s3) [place]  [right=\eodisty of s2, label=above:$s_3$] {};
    \node (t3) [transition] [below=\eodisty of s3, label=right:$a$] {};
    \node (s6) [place]  [below=\eodisty of t3, label=below:$s_6$] {};
    
    \draw  [->] (s1) to (t1);
    \draw  [->] (t1) to (s4);
    \draw  [->] (s2) to (t2);
    \draw  [->] (t2) to (s5);
    \draw  [->] (s3) to (t3);
    \draw  [->] (t3) to (s6);    
    
    \draw [-o] (s1) to (t2);
    \draw [-o] (s3) to (t2);
    
    \node (s1p) [place] [right={4.5cm} of s1, label=above:$s_1'$] {};
    \node (t1p) [transition] [below=\eodisty of s1p, label=left:$a$] {};
    \node (s4p) [place]  [below right=\eodist of t1p, label=below:$s_4'$] {};
    
    \node (s2p) [place]  [right=\eodisty of s1p, label=above:$s_2'$] {};
    \node (t2p) [transition] [below=\eodisty of s2p, label=right:$b$] {};
    
    \draw  [->] (s1p) to (t1p);
    \draw  [->] (t1p) to (s4p);
    \draw  [->] (s2p) to (t2p);
    \draw  [->] (t2p) to (s4p);
    
    \draw [-o] (s1p) to (t2p);
}

\end{tikzpicture}
    
    \caption{Two PTI nets, whose transitions are labeled either by $a$ or by $b$. 
    }
    \label{fig:2}
\end{figure}
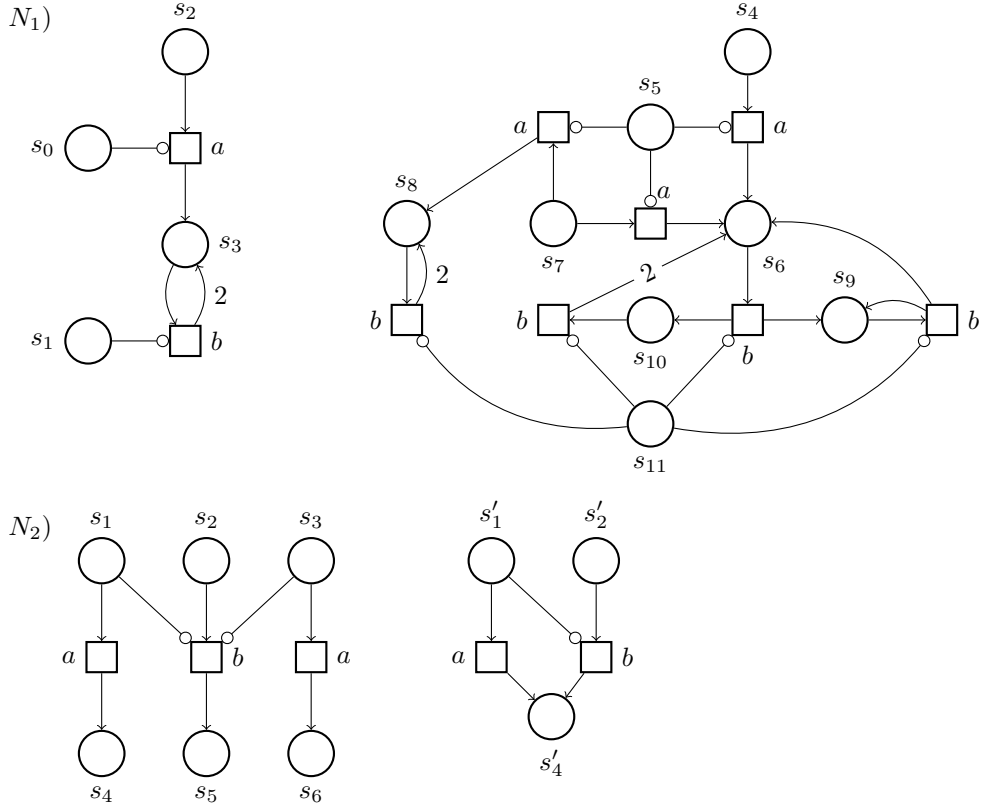

\begin{example}
Consider the PTI net $N_1$ in \cref{fig:2}. Not only the loop labeled by $b$ on the left is
unwound on the right, but also the $a$-labeled transition on the left is 
replicated three times on the right.
The relation 
\begin{center}
$R = \{
(s_0, s_5),$ $ (s_1, s_{11}), (s_2, s_4), (s_2, s_7),
(s_3, s_6), (s_3, s_8),$ $(s_3, s_9), (s_3, s_{10})
\}$    
\end{center}
is a pti-place bisimulation and so, e.g.,
$2 \cdot s_2 \oplus s_3 \sim_p s_4 \oplus s_7 \oplus s_9$.

Now consider the PTI net $N_2$ in \cref{fig:2}. 
In this case, the $b$-labeled transition on the left can be matched by the $b$-labeled transition 
on the right, even if their inhibiting set differ in size, because
both $(s_1, s_1')$ and $(s_3, s_1')$ are in the following bisimulation.
Indeed, the relation 
\begin{center}
 $R' = \{
(s_1, s_1', (s_2, s_2'), (s_3, s_1'), 
(s_4, s_4'), (s_5, s_4'), (s_6, s_4')
\}$    
\end{center}
is a pti-place bisimulation and so, e.g.,
$s_1 \oplus s_3 \oplus 2 \cdot s_2 \oplus s_5 \sim_p 2 \cdot s_1' \oplus 2 \cdot s_2' \oplus s_4'$.

\end{example}

\begin{figure}[t]
    \centering
    \begin{tikzpicture}[
        every place/.style={draw,thick,inner sep=0pt,minimum size=6mm},
        every transition/.style={draw,thick,inner sep=0pt,minimum size=4mm},
        bend angle=30,
        pre/.style={<-,shorten <=1pt,>=stealth,semithick},
        post/.style={->,shorten >=1pt,>=stealth,semithick}
    ]
    \def\eofigdist{3.3cm}
    \def\eodist{0.5cm}
    \def\eodisty{0.75cm}

    \node (UPC) [label=$UPAC$]{};
    \node (p) [place] [below={0.7cm} of UPC, label=above:$p$]{};
    \node (ta) [transition] [below left = {1.5cm} of p, label=left:$a$] {};
    \node (wa) [place] [below = {1cm} of ta, label=left:$w_a$] {};
    \node (da) [transition] [below = {1cm} of wa, label=left:$o_a$] {};
    
    \node (tan) [transition] [below = {1cm} of p, label=right:$a$] {};
    \node (wan) [place] [below = {1cm} of tan, label=right:$w'_a$] {};
    \node (dan) [transition] [below = {1cm} of wan, label=right:$o_a$] {};
    
    \node (s) [place] [below of = dan, label=left:$s$] {};
    \node (t) [transition] [below of = s, label=left:$d$] {};
    \node (c) [place] [below of = t, label=below:$c$] {};

    \node (tb) [transition] [below right = {1.5cm} of p, label=right:$b$] {};
    \node (wb) [place] [below = {1cm} of tb, label=right:$w_b$] {};
    \node (db) [transition] [below = {1cm} of wb, label=right:$o_b$] {};

    \draw  [->] (p) to (ta);
    \draw  [->] (ta) to (wa);
    \draw  [->] [bend left] (ta) to (p);
    \draw  [->] (wa) to (da);
    \draw  [->] (da) to (s);
    \draw  [->] (s) to (t);
    \draw  [->] (t) to (c);
    \draw  [->] [bend left] (c) to (da);
    \draw  [-o] (wa) to (tan);
    \draw  [-o] [bend left](wa) to (ta);

    \draw  [->] (p) to (tan);
    \draw  [->] (tan) to (wan);
    \draw  [->] [bend right] (tan) to (p);
    \draw  [->] (wan) to (dan);
    \draw  [->] (dan) to (s);
    \draw  [->] [bend right] (c) to (dan);
    \draw  [-o] (wan) to (ta);
    \draw  [-o] [bend right] (wan) to (tan);
    \draw  [-o] (wan) to (tb);
    \draw  [-o] (wan) to (db);

    \draw  [-o] (wa) to (tb);
    \draw  [-o] (wa) to (db);
    \draw  [->] [bend right] (tb) to (p);
    \draw  [->] (p) to (tb);
    \draw  [->] (tb) to (wb);
    \draw  [->] (wb) to (db);
    \draw  [->] (db) to (s);
    \draw  [->] [bend right] (c) to (db);
    
    \node (UPPC) [right={5.5cm} of UPC, label=$UPBC$]{};
    \node (pp) [place] [below={0.7cm} of UPPC, label=above:$\overline{p}$]{};
    \node (tap) [transition] [below left = {1.5cm} of pp, label=left:${a}$] {};
    \node (wap) [place] [below = {1cm}  of tap, label=left:$\overline{w}_a$] {};
    \node (dap) [transition] [below = {1cm}  of wap, label=left:$ o_a$] {};
    
    \node (tbp) [transition] [below = {1cm} of pp, label=right:$ b$] {};
    \node (wbp) [place] [below = {1cm}  of tbp, label=right:$\overline w_b$] {};
    \node (dbp) [transition] [below = {1cm}  of wbp, label=right:$ o_b$] {};

    \node (tbs) [transition] [below right= {1.5cm} of pp, label=right:$ b$] {};
    \node (wbs) [place] [below  = {1cm} of tbs, label=right:$\overline w'_b$] {};
    \node (dbs) [transition] [below  = {1cm} of wbs, label=right:$ o_b$] {};
    
    \node (sp) [place] [below of = dbp, label=left:$\overline s$] {};
    \node (tp) [transition] [below of = sp, label=left:$ d$] {};
    \node (cp) [place] [below of = tp, label=below:$\overline c$] {};

    \draw  [->] (pp) to (tap);
    \draw  [->] (tap) to (wap);
    \draw  [->] [bend left] (tap) to (pp);
    \draw  [->] (wap) to (dap);
    \draw  [->] (dap) to (sp);
    \draw  [->] (sp) to (tp);
    \draw  [->] (tp) to (cp);
    \draw  [->] [bend left] (cp) to (dap);
    \draw  [-o] [bend left] (wap) to (tap);

    \draw  [-o] (wap) to (tbp);
    \draw  [-o] (wap) to (dbp);
    \draw  [->] [bend right] (tbp) to (pp);
    \draw  [->] (pp) to (tbp);
    \draw  [->] (tbp) to (wbp);
    \draw  [->] (wbp) to (dbp);
    \draw  [->] (dbp) to (sp);
    \draw  [->] [bend right] (cp) to (dbp);

    \draw  [-o] (wap) to (tbs);
    \draw  [-o] (wap) to (dbs);
    \draw  [->] [bend right] (tbs) to (pp);
    \draw  [->] (pp) to (tbs);
    \draw  [->] (tbs) to (wbs);
    \draw  [->] (wbs) to (dbs);
    \draw  [->] (dbs) to (sp);
    \draw  [->] [bend right] (cp) to (dbs);

\end{tikzpicture}
    
    \caption{A PTI net representing two unbounded producers/consumers with priority. 
    For simplicity, we display the labels of transitions instead of their names.
    }
    \label{fig:upc}
\end{figure}
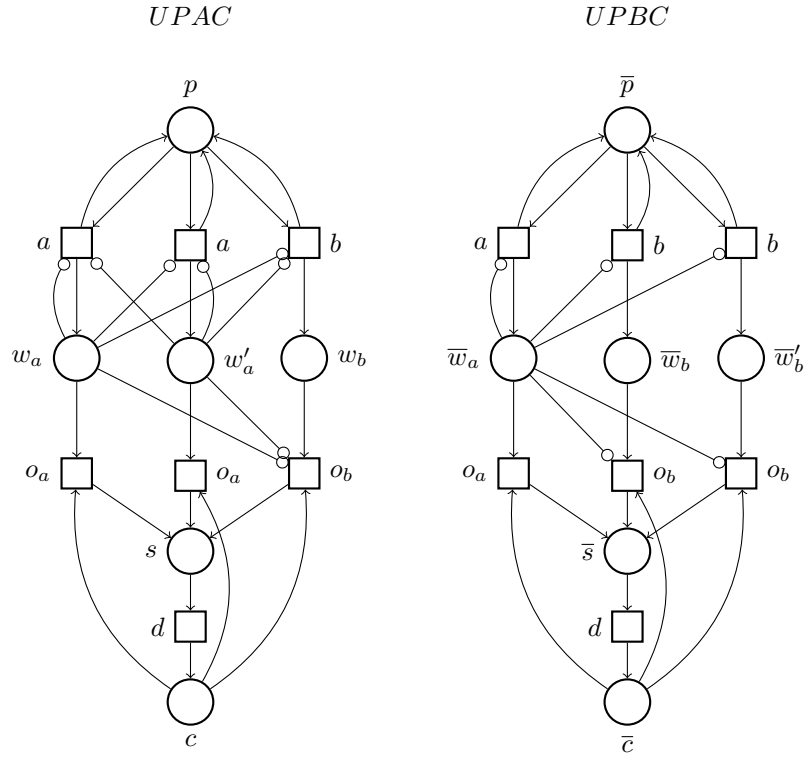

\begin{example}
Consider the PTI net in \cref{fig:upc}, depicting two models of unbounded producer-consumer with priority.

In the left part, denoted as $UPAC$ for readability, 
the producer $p$ can generate two products of type $a$ and $b$ and stock them
in $w_a$ and $w_b$ (for "warehouse") respectively. Transitions $o_a$ and $o_b$ model the order
of a client $c$ from the warehouse, which may then be shipped (place $s$) and delivered
(transition $d$).
Product $a$ has priority both in the production and ordering phases, and this is modelled by two 
inhibiting arcs between $w_a$ and $b$ and $o_b$. Roughly speaking, if there is an $a$ in the warehouse,
then no $b$ can be produced or ordered. Moreover, only one product of type $a$ can be stored in 
either $w_a$ or $w'_a$,
as the inhibiting arcs between a warehouse for $a$ and the other $a$-transition 
do not allow to perform the latter until
the former has been freed by the execution of transition $o_a$.

In the right part, denoted as $UPBC$, we duplicate the production and ordering phases of product $b$, 
and remove one of the two lines of product $a$.
The behavior of the system remains the same, and this is proved by the pti-place bisimulation
\begin{center}
$R = \{ (p, \overline p), (w_a, \overline w_a),
(w_a', \overline w_a),
(w_b, \overline w_b),  (w_b, \overline w_b'),
(s, \overline s),
(c, \overline c)\}$.
\end{center}
\end{example}

We now prove that $\sim_p$ is an equivalence relation.

\begin{proposition}\label{pti-place-bis-eq}
For each PTI net $N = (S, A, T, I)$, relation $\sim_p \; \subseteq  \mathcal{M}(S) \times  \mathcal{M}(S)$ is an equivalence relation.
\begin{proof}
Direct consequence of the fact that
for each PTI net $N = (S, A, T, I)$, the following hold:
\begin{enumerate}
\item The identity relation ${\mathcal I}_S = \{ (s, s) \mid s \in S \}$ is a pti-place bisimulation;
\item the inverse relation $R^{-1} = \{ (s', s) \mid (s, s') \in R\}$ of a pti-place bisimulation $R$ is a pti-place bisimulation;
\item the relational composition $R_1 \circ R_2 = \{ (s, s'') \mid $ $\exists s'. (s, s') \in R_1 \wedge (s', s'') \in R_2 \}$ of
two pti-place bisimulations $R_1$ and $R_2$ is a pti-place bisimulation.
\end{enumerate}
See \cref{pti_place_bis_equivalence_sec} for details.
\end{proof}
\end{proposition}

\noindent
By \cref{def-pti-place-bis}, pti-place bisimilarity can be defined in the following way:

$\sim_p = \bigcup \{ R^\oplus \mid R \mbox{ is a pti-place bisimulation}\}.$

\noindent
By monotonicity of the additive closure (\cref{add-prop1}(2)), if $R_1 \subseteq R_2$, then
$R_1^\oplus \subseteq R_2^\oplus$. 
Hence, we can restrict our attention to maximal pti-place bisimulations only:

$\sim_p = \bigcup \{ R^\oplus \mid R \mbox{ is a {\em maximal} pti-place bisimulation}\}.$

\noindent
However, it is not true that 

$\sim_p = (\bigcup \{ R \mid R \mbox{ is a {\em maximal} pti-place bisimulation}\})^\oplus$

\noindent 
because the union of pti-place bisimulations may be not a pti-place bisimulation 
(as already observed  for place bisimulation in \cite{ABS91,Gor21}), 
so that its definition is not coinductive.

\begin{figure}[t]
    \centering
    
    \begin{tikzpicture}[
        every place/.style={draw,thick,inner sep=0pt,minimum size=6mm},
        every transition/.style={draw,thick,inner sep=0pt,minimum size=4mm},
        bend angle=30,
        pre/.style={<-,shorten <=1pt,>=stealth,semithick},
        post/.style={->,shorten >=1pt,>=stealth,semithick}    
    ]
    
    \node (s2) [place]  [label=below:$s_2$] {};
    \node (t2) [transition] [below right of = s2,label=above:$a$] {};
    \node (s3) [place]  [above right of = t2, label=below:$s_3$] {};
    \node (t1) [transition] [left of = s2,label=above:$a$] {};
    \node (s1) [place]  [left of = t1, label=above:$s_1$] {};
    \node (t3) [transition] [right of = s3,label=above:$a$] {};
    \node (s4) [place]  [right of = t3, label=above:$s_4$] {};
    \node (s5) [place]  [below of = t2, label=below:$s_5$] {};
    
    \draw [->] (s2) to (t2);
    \draw [->] (s3) to (t2);
    \draw [->] (s2) to (t1);
    \draw [->] (t1) to (s1);
    \draw [->] (s3) to (t3);
    \draw [->] (t3) to (s4);
    \draw [->] (t2) to (s5);
    \draw [-o, bend right] (s3) to (t1);
    \draw [-o, bend left] (s2) to (t3);
    \end{tikzpicture}

    \caption{A PTI net.}
    \label{fig:pti_not_coinductive}
\end{figure}

\begin{example}
Consider the net in \cref{fig:pti_not_coinductive}, whose transitions are $t_1 = (s_2, s_3) \deriv{a} s_1$,
$t_2  = (s_2 \oplus s_3, \theta) \deriv{a} s_5$ and $t_3 = (s_3, s_2) \deriv{a} s_4$.
Clearly, $R_1$ and $R_2$, defined as follows, are both maximal pti-place
bisimulations.
\begin{center}
$R_1 = \{(s_2, s_2), (s_3, s_3))\} \cup (\{s_1, s_4, s_5\} \times \{s_1, s_4, s_5\})$\\
$R_2 = \{(s_2, s_3), (s_3, s_2)\}\cup (\{s_1, s_4, s_5\} \times \{s_1, s_4, s_5\})$
\end{center}

Note that the union $R = R_1 \cup R_2$ is not a pti-place bisimulation
as, for example,
$(2 \cdot s_2, s_2 \oplus s_3) \in R^\oplus$, but the pti-place bisimulation conditions are not satisfied.
Indeed, if $2 \cdot s_2$ moves  first by
$2 \cdot s_2 \trns{t_1} s_1 \oplus s_2$, then
$s_2 \oplus s_3$ can only try to respond with
$s_2 \oplus s_3 \trns{t_2} s_5$ since
$t_1$ and $t_3$ are inhibited.
However, this is not possible because we have that $(\pre t_1, \pre t_2) \not\in R^\oplus$, 
and, even worse, $(s_2, \theta) \not\in R^\oplus$.
\end{example}

\subsection{Pti-place bisimilarity is finer than causal-net bisimilarity}

\begin{theorem}\label{pti-bis-finer-cn-bis}
{\bf (Pti-place bisimilarity implies causal-net bisimilarity)}
Let $N=(S,A,T,I)$ be a PTI net and $m_1, m_2$ two of its markings.
If $m_1 \sim_{p} m_2$, then $m_1 \sim_{cn} m_2$.

\begin{proof}
See \cref{pti-cn-bis-th}.
\end{proof}
\end{theorem}

There are at least the following three important technical differences between 
causal-net bisimilarity and pti-place bisimilarity. 
\begin{enumerate}
    \item A causal-net bisimulation is a very complex relation --
        composed of cumbersome triples of the form $(\rho_1, C, \rho_2)$ --
        that must contain infinitely many triples if the net system offers a never-ending behavior. 
        On the contrary,
        a pti-place bisimulation is always a very simple finite relation over the finite set $S$ of places.
        
    \item A causal net bisimulation  proving that $m_1 \sim_{cn} m_2 $
            is a relation specifically designed for showing that  $m_1$ and $m_2$ 
            generate the same causal nets, step by step. If we want to prove that,
            e.g., $n \cdot m_1$ and $n \cdot m_2$ are causal-net bisimilar (which 
            may not hold!), we have to construct a new causal-net bisimulation to this aim. 
            Instead, a pti-place bisimulation $R$ 
            relates those places which are considered equivalent under all the possible 
            $R$-related contexts. 
            Hence, if $R$ justifies that  $m_1 \sim_{p} m_2 $ 
            as $(m_1, m_2) \in R^\oplus$, then
            for sure the same $R$ justifies that $n \cdot m_1$ and $n \cdot m_2$ are 
            pti-place bisimilar, as also 
            $(n \cdot m_1, n \cdot m_2) \in R^\oplus$.
            
    \item Finally, while pti-place bisimilarity is decidable (see the next section), 
        it is not known whether causal-net bisimilarity is decidable on finite PTI nets.\footnote{Esparza observed \cite{Esp98}
        that, for finite P/T nets with at least two unbounded places, all the behavioral relations ranging from interleaving bisimilarity
        to fully-concurrent bisimilarity \cite{BDKP91} are undecidable. Even if his proof does not apply to causal-net bisimilarity, 
        we conjecture that this equivalence is undecidable as well.}
        
\end{enumerate}
However, these technical advantages of pti-place bisimilarity over causal-net bisimilarity are 
balanced by an increased discriminating power of the former over the latter, that, in some cases, 
might appear even excessive, as the following intriguing example shows.

\begin{figure}
    \centering
    \begin{tikzpicture}[
        every place/.style={draw,thick,inner sep=0pt,minimum size=6mm},
        every transition/.style={draw,thick,inner sep=0pt,minimum size=4mm},
        bend angle=30,
        pre/.style={<-,shorten <=1pt,>=stealth,semithick},
        post/.style={->,shorten >=1pt,>=stealth,semithick}
    ]
    \def\eofigdist{3.3cm}
    \def\eodist{0.35cm}
    \def\eodisty{0.55cm}

\node (s1) [place]  [label=above:$s_1$] {};
\node (t1) [transition] [below=\eodisty of s1, label=left:$a$] {};
\node (s3) [place]  [below right=\eodisty of t1, label=right:$s_3$] {};
\node (s2) [place]  [below left=\eodisty of t1, label=left:$s_2$] {};
\node (t2) [transition] [below left=\eodisty of s3, label=left:$b$] {};
\node (s4) [place]  [right=\eodisty of t2, label=below:$s_4$] {};

\draw  [->] (s1) to (t1);
\draw  [->] (t1) to (s2);
\draw  [->] (t1) to (s3);
\draw  [-o] [bend right] (s3) to (t1);
\draw  [->] (s2) to (t2);
\draw  [->] (s3) to (t2);
\draw  [->] (t2) to (s4);

\node (s5) [place]  [right={4cm} of s1, label=above:$s_1'$] {};
\node (t3) [transition] [below=\eodisty of s5, label=left:$a$] {};
\node (s6) [place]  [below=\eodisty of t3, label=right:$s_2'$] {};
\node (t4) [transition] [below=\eodisty of s6, label=left:$b$] {};
\node (s7) [place]  [right=\eodisty of t4, label=right:$s_4'$] {};
\draw  [->] (s5) to (t3);
\draw  [->] (t3) to node[auto,swap] {2} (s6);
\draw  [-o] [bend right] (s6) to (t3);
\draw  [->] (s6) to node[auto,swap] {2} (t4);
\draw  [->] (t4) to (s7);

\end{tikzpicture}
    
    \caption{Two PTI nets. }
    \label{fig:3}
\end{figure}

\begin{example}\label{ex-strano}
Consider the net in \cref{fig:3}. 
First of all, note that  $s_2 \sim_{cn} s_2'$, because both are stuck markings.
However, we have that $2 \cdot s_2 \nsim_{cn} 2 \cdot s_2'$ because 
$2 \cdot s_2$ is stuck, while $2 \cdot s_2'$ can perform $b$. 
This observation is enough to conclude that $s_2 \nsim_p s_2'$, 
because a pti-place bisimulation $R$ relates places that 
are equivalent under any $R$-related context: 
if $(s_2, s_2') \in R$ then $(2 \cdot s_2, 2 \cdot s_2') \in R^\oplus$, 
but these two markings do not satisfy the pti-place bisimulation conditions, 
so $R$ is not a pti-place bisimulation.

Nonetheless, it is interesting to observe that $s_1 \sim_{cn} s_1'$, because they generate the same 
causal PTI nets, step by step; moreover,  even for any $n \geq 1$ we have
$n \cdot s_1 \sim_{cn} n \cdot s_1'$. However, $s_1 \nsim_p s_1'$ 
because it is not possible to build a pti-place bisimulation $R$ containing the pair $(s_1, s_1')$.
The problem is that it would be necessary to include, into the candidate pti-place relation $R$, 
also the pair $(s_2, s_2')$, which is not a pti-place bisimulation pair, as discussed above.
Therefore, no pti-place bisimulation $R$ can relate $s_1$ and $s_1'$.
\end{example}

\section{Pti-place bisimilarity is decidable}\label{decid-iplace-sec}
In order to prove that $\sim_p$ is decidable, we first need
a technical lemma which states that it is decidable
to check whether a place relation $R \subseteq S \times S$ 
is a pti-place bisimulation.

\begin{lemma}\label{r-pti-place-bis-decid}
Given a finite PTI net $N = (S, A, T, I )$ and a
place relation $R \subseteq S \times S$,
it is decidable whether $R$ is a pti-place bisimulation.
\begin{proof}
It is enough to check two finite conditions on transitions and places of the net; full detail in \cref{r_pti_place_bis_decid_sec}.
\end{proof}

\end{lemma}

\begin{theorem} \label{pti-place-decidable}
{\bf (Pti-place bisimilarity is decidable)}
Given a PTI net $N = (S, A, T, I )$, 
for each pair of markings $m_1$ and $m_2$, 
it is decidable whether $m_1 \sim_p m_2$.

\begin{proof}
If $|m_1| \neq |m_2|$, then $m_1 \nsim_p m_2$ by \cref{fin-k-add}. Otherwise, we can assume that $|m_1| = k = |m_2|$.
As $|S| = n$, the set of all the place relations over $S$ is of size $2^n$. 
Let us list such relations as: 
$R_1, R_2, \ldots, R_{2^n}$.
Hence, for $i = 1, \ldots, 2^n$, by \cref{r-pti-place-bis-decid} 
we can decide whether the place relation $R_i$ is 
a pti-place bisimulation and, in such a case,
we can check whether $(m_1, m_2) \in R_i^\oplus$ in $O(k^2 \sqrt{k})$ time. 
As soon as we have found a pti-place bisimulation $R_i$ such that 
$(m_1, m_2) \in R_i^\oplus$,
we stop concluding that $m_1 \sim_p m_2$. 
If none of the $R_i$ is a pti-place bisimulation such that 
$(m_1, m_2) \in R_i^\oplus$, then
we can conclude that $m_1 \nsim_p m_2$. 
Since this procedure might scan all place relations, the worst-case complexity of the algorithm is exponential in the number of places $n$.
\end{proof}
\end{theorem}

\section{Conclusion}\label{conc-sec}

Pti-place bisimilarity is the only decidable behavioral equivalence for finite PTI nets, which constitute 
a powerful, Turing-complete 
distributed model of computation, widely used in theory and applications of concurrency 
(e.g., \cite{ager-pti,ajmone,BP99,tcsinib,BG09,Hack,Koutny,Pet81}).
Thus, it is the only equivalence for which it is possible (at least, in principle) 
to verify algorithmically the (causality-preserving) correctness of an implementation by exhibiting a pti-place bisimulation between its
specification and implementation. It is also sensible, because it respects the causal behavior of PTI nets, since it is finer than
causal-net bisimilarity. Of course, pti-place bisimilarity is a rather discriminating behavioral equivalence, 
as illustrated in Example \ref{ex-strano},
and a proper evaluation of its usefulness on real case studies is left for future research.

In our interpretation, (pti-)place bisimilarity is 
an attempt of giving semantics to {\em unmarked}, rather than marked, nets,
shifting the focus from the usually undecidable question {\em When are two markings equivalent?} to the decidable (but more restrictive) question {\em When are two places equivalent?}
A possible answer to the latter question may be: two places are equivalent if, 
whenever the same number of tokens are put on these two places,
the behavior of the marked nets is the same. If we reinterpret Example \ref{ex-strano} in this perspective, we clearly see that
place $s_2$ and place $s_2'$ cannot be considered as equivalent because, even if the marking $s_2$ and $s_2'$ are equivalent
(as they are both stuck), the marking $2 \cdot s_2$ is not equivalent
to the marking $2 \cdot s_2'$ (as only the latter can move).
More specifically, a (pti-)place bisimulation $R$ considers two places $s_1$ and $s_2$ as equivalent if $(s_1, s_2) \in R$, as,  by definition 
of (pti-)place bisimulation, they must behave the same in any $R$-related context. 

The decidability result for pti-place bisimilarity is based on the fact that 
the net model is finite, even if the associated reachability graph may be unboundedly large or even infinite: indeed, 
one can decide pti-place bisimilarity simply checking a large, but finite, number of conditions on the shape of the 
finite net, rather than inspecting its (possibly, infinitely many) reachable markings.

Turing completeness is achieved in PTI nets by means of their ability to test for zero. 
Other Turing-complete models of computation may exploit different
mechanisms to this aim. For instance, in the $\pi$-calculus \cite{MPW,SW} Turing completeness is achieved by means of the ability to generate unboundedly new names (by means of the interplay between recursion and the restriction operator), but this feature is not describable
by means of a finite net model \cite{BG09,MG09}.
For this reason, we think it is hard to find a sensible, decidable behavioral equivalence for the whole $\pi$-calculus.

To the best of our knowledge, this is the second paper proving the decidability of a behavioral equivalence for a Turing-complete 
formalism. In fact, in \cite{lanese} it is proved that 
(interleaving) bisimilarity is decidable for a small process calculus, called HOcore, with higher-order communication (but without restriction), 
that is, nonetheless, Turing-complete.

Future work will be devoted to see whether the pti-place bisimulation idea can be extended to other, 
possibly even larger classes of nets, such as 
{\em lending} Petri nets \cite{P15}, where transitions are allowed to consume tokens from a place
even if it does not contain enough tokens, thus enabling negative-valued markings.


\appendix

\section{Properties of pti-place bisimilarity}

\subsection{Pti-place bisimilarity is an equivalence}\label{pti_place_bis_equivalence_sec}
\begin{proposition}\label{pti-prop-bis}
For each PTI net $N = (S, A, T, I)$, the following hold:
\begin{enumerate}
\item The identity relation ${\mathcal I}_S = \{ (s, s) \mid s \in S \}$ is a pti-place bisimulation;
\item the inverse relation $R^{-1} = \{ (s', s) \mid (s, s') \in R\}$ of a pti-place bisimulation $R$ is a pti-place bisimulation;
\item the relational composition $R_1 \circ R_2 = \{ (s, s'') \mid $ $\exists s'. (s, s') \in R_1 \wedge (s', s'') \in R_2 \}$ of
two pti-place bisimulations $R_1$ and $R_2$ is a pti-place bisimulation.
\end{enumerate}

\begin{proof}
The proof is almost standard, due to \cref{add-prop2}.

(1) ${\mathcal I}_S$ is a pti-place bisimulation as for each $(m, m) \in {\mathcal I}_S^\oplus$
whatever transition $t$ the left (or right) marking $m$ performs a transition (say,  $m[t\rangle m'$), 
the right (or left) 
instance of $m$
in the pair does exactly the same transition 
$m[t\rangle m'$ and, of course, $(\pre{t}, \pre{t}) \in {\mathcal I}_S^\oplus$, 
          $(\post{t}, \post{t}) \in {\mathcal I}_S^\oplus$, $l(t) = l(t)$, $(m \ominus \pre{t}, m \ominus \pre{t}) \in {\mathcal I}_S^\oplus$, by \cref{add-prop2}(2),
          and, also,  $\forall s \in S. (s, s) \in {\mathcal I}_S \Rightarrow (s \in \prei{t} \Leftrightarrow$ $ s \in \prei{t})$, 
          as required by the pti-place bisimulation definition.

(2) Suppose $(m_2, m_1) \in (R^{-1})^\oplus$ and $m_2[t_2\rangle m_2'$. By \cref{add-prop2}(3)
$(m_2, m_1) \in (R^\oplus)^{-1}$ and so $(m_1, m_2) \in R^\oplus$. Since $R$ is a pti-place bisimulation, item 2 of the 
bisimulation game ensures that there exist $t_1$ and $m_1'$
such that $m_1 [t_1\rangle m_1'$, with $(\pre{t_1}, \pre{t_2}) \in R^\oplus$,
$l(t_1) = l(t_2)$, $(\post{t_1}, \post{t_2}) \in R^\oplus$
and $(m_1 \ominus \pre{t_1}, m_2 \ominus \pre{t_2}) \in R^\oplus$;
moreover, $\forall s, s' \in S. (s, s') \in R \Rightarrow (s \in \prei{t_1} \Leftrightarrow s' \in \prei{t_2})$.
Summing up, if $(m_2, m_1) \in (R^{-1})^\oplus$,
to the move $m_2[t_2\rangle m_2'$, $m_1$ replies with the move $m_1 [t_1\rangle m_1'$, such that 
(by \cref{add-prop2}(3))
$(\pre{t_2}, \pre{t_1}) \in (R^{-1})^\oplus$, 
$l(t_2) = l(t_1)$, $(\post{t_2}, \post{t_1}) \in (R^{-1})^\oplus$, $(m_2 \ominus \pre{t_2}, m_1 \ominus \pre{t_1}) \in (R^{-1})^\oplus$ and, moreover, 
$\forall s, s' \in S. (s', s) \in R^{-1} \Rightarrow (s' \in \prei{t_2} \Leftrightarrow s \in \prei{t_1})$, as required. 
The case when $m_1$ 
moves first is symmetric and thus omitted.

(3)  Suppose $(m, m'') \in (R_1 \circ R_2)^\oplus$ and $m [t_1\rangle m_1$.
By \cref{add-prop2}(4), we have that 
$(m, m'') \in R_1^\oplus \circ R_2^\oplus$, and so there exists $m'$ such that $(m, m') \in R_1^\oplus$
and $(m', m'') \in R_2^\oplus$.
As $(m, m') \in R_1^\oplus$ and $R_1$ is a pti-place bisimulation, 
if $m [t_1\rangle m_1$, then there exist $t_2$ and $m_2$ such that  
$m' [t_2\rangle m_2$ with $(\pre{t_1}, \pre{t_2}) \in R_1^\oplus$,
$l(t_1) = l(t_2)$, $(\post{t_1}, \post{t_2}) \in R_1^\oplus$
and $(m \ominus \pre{t_1}, m' \ominus \pre{t_2}) \in R_1^\oplus$; moreover, $\forall s, s' \in S. (s, s') \in R_1 \Rightarrow (s \in \prei{t_1} \Leftrightarrow s' \in \prei{t_2})$.
But as $(m', m'') \in R_2^\oplus$ and $R_2$ is a pti-place bisimulation, we have also that there exist $t_3$ and $m_3$ 
such that  $m'' [t_3 \rangle m_3$ 
with $(\pre{t_2}, \pre{t_3}) \in R_2^\oplus$, 
$l(t_2) = l(t_3)$, $(\post{t_2}, \post{t_3}) \in R_2^\oplus$
and $(m' \ominus \pre{t_2}, m'' \ominus \pre{t_3}) \in R_2^\oplus$; moreover, $\forall s', s'' \in S. (s', s'') \in R_2 
\Rightarrow (s' \in \prei{t_2} \Leftrightarrow s'' \in \prei{t_3})$.
Summing up, for $(m, m'') \in (R_1 \circ R_2)^\oplus$, if $m [t_1 \rangle m_1$, then there exist $t_3$ and $m_3$ such that 
$m'' [t_3 \rangle m_3$ and (by \cref{add-prop2}(4)) $(\pre{t_1}, \pre{t_3}) \in (R_1 \circ R_2)^\oplus$, 
$l(t_1) = l(t_3)$, $(\post{t_1}, \post{t_3}) \in (R_1 \circ R_2)^\oplus$
and $(m \ominus \pre{t_1}, m'' \ominus \pre{t_3}) \in (R_1 \circ R_2)^\oplus$; moreover, $\forall s, s'' \in S. (s, s'') \in 
R_1 \circ R_2 \Rightarrow (s \in \prei{t_1} \Leftrightarrow s'' \in \prei{t_3})$,
as required. The case when $m''$ 
moves first is symmetric and so omitted.
\end{proof}
\end{proposition}

\begin{proposition}
For each PTI net $N = (S, A, T, I)$, relation $\sim_p \; \subseteq  \mathcal{M}(S) \times  \mathcal{M}(S)$ is an equivalence relation.
\begin{proof}
Direct consequence of \cref{pti-prop-bis}.
\end{proof}
\end{proposition}

\subsection{Pti-place bisimilarity is finer than causal-net bisimilarity}\label{pti-cn-bis-th}

\begin{theorem}
{\bf (Pti-place bisimilarity implies causal-net bisimilarity)}
Let $N=(S,A,T,I)$ be a PTI net and $m_1, m_2$ two of its markings.
If $m_1 \sim_{p} m_2$, then $m_1 \sim_{cn} m_2$.

\begin{proof}
If $m_1 \sim_{p} m_2$, then there exists a pti-bisimulation $R_1$ such that
$(m_1, m_2) \in R_1^\oplus$.
Let us consider 
    \begin{equation*} \label{R2}
        \begin{split}
        R_2 \overset{def}{=} \lbrace (\rho_1, C, \rho_2) | & (C, \rho_1) \text{ is a PTI process of $N(m_{1})$ and} \\
        &(C, \rho_2) \text{ is a PTI process of $N(m_{2})$ and} \\
        & \forall b \in B \; (\rho_1(b), \rho_2(b)) \in R_1
         \rbrace .
        \end{split}
    \end{equation*}

\noindent
We want to prove that $R_2$ is a causal-net bisimulation.
First of all, consider a triple of the form $(\rho_1^0, C^0, \rho_2^0)$,
where $C^0$ is the causal PTI net without events and $\rho_1^0, \rho_2^0$ are
such that  $\rho_i^0(Min(C^0)) = \rho_i^0(Max(C^0)) = \rho_i^0(B^0) = m_i$ for $i= 1, 2$,
and $(\rho_1^0(b), \rho_2^0(b)) \in R_1$ for all $b \in B^0$.
Then $(\rho_1^0, C^0, \rho_2^0)$ must belong to $R_2$,
because $(C^0, \rho_i^0)$ is a process of $N(m_i)$, for $i=1, 2$ and, 
by hypothesis, $(m_1, m_2) \in R_1^\oplus$. 
Hence, if $R_2$ is a causal-net bisimulation, then the triple 
$(\rho_1^0, C^0, \rho_2^0) \in R_2$ ensures that $m_1 \sim_{cn} m_2$. 

Assume $\cnt \in R_2$. 
In order for $R_2$ to be a cn-bisimulation,
we must prove that
\begin{romanenumerate}
\item
$\forall t_1, C', \rho_1'$ such that $(C, \rho_1) \deriv{e} (C', \rho_1')$,
where $\rho_1'(e) = t_1$,
$\exists t_2, \rho_2'$ such that\\
$(C, \rho_2) \deriv{e} (C', \rho_2')$,
where $\rho_2'(e) = t_2$, and
$(\rho'_1, C', \rho'_2) \in R_2$;

\item symmetrical, if $(C, \rho_2)$ moves first.
\end{romanenumerate}
Assume $(C, \rho_1) \deriv{e} (C', \rho_1')$ with $\rho_1'(e) = t_1$.
Since $\cnt \in R_2$, for all $b \in Max(C)$ we have $(\rho_1(b), \rho_2(b)) \in R_1$
and therefore $(\rho_1(Max(C)), \rho_2(Max(C))) \in R_1^\oplus$.
Since $\rho_1(Max(C)) \\ \trns{t_1} \rho_1'(Max(C'))$ and $R_1$ is a pti-place bisimulation,
there exist $t_2, m_2$ such that 
$\rho_2(Max(C)) \\ \trns{t_2} m_2$ with
$(\pre{t_1}, \pre{t_2}) \in R_1^\oplus$,
$l(t_1) = l(t_2)$,
$(\post{t_1}, \post{t_2}) \in R_1^\oplus$,\\ $(\rho_1(Max(C)) \ominus \pre{\rho_1'(e)},$ $\rho_2(Max(C)) \ominus \pre{\rho_2'(e)}) \in R_1^\oplus$ and, moreover, 
$\forall s, s' \in S. (s, s') \in R_1 \Rightarrow (s \in \prei{t_1} \Leftrightarrow s' \in \prei{t_2})$.
Note that, since $(\post{t_1}, \post{t_2}) \in R_1^\oplus$ and $(\rho_1(Max(C)) \ominus \pre{\rho_1'(e)}, \rho_2(Max(C)) \ominus \pre{\rho_2'(e)}) \in R_1^\oplus$, by additivity of additive closure (cf. \cref{add-prop1}), 
$(\rho_1(Max(C)) \ominus \pre{\rho_1'(e)} \oplus \post{t_1}, \rho_2(Max(C)) \ominus \pre{\rho_2'(e)} \oplus \post{t_2}) \in R_1^\oplus$, i.e. $(\rho_1'(Max(C')), m_2) \in R^\oplus$.

Therefore, since $t_1$ and $t_2$ have the same pre-sets/post-sets up to $R_1$, 
it is possible to derive $(C, \rho_2) \deriv{e} (C'', \rho_2')$,
where $\rho_2'$ is such that $\rho_2'(e) = t_2$ and $(\rho_1'(b), \rho_2'(b)) \in R_1$ for each $b \in \post{e}$ (which is really possible because $(\post{t_1}, \post{t_2}) \in R_1^\oplus$).
Now we prove that $C' = C''$. The underlying P/T parts of $C'$ and $C''$ are obviously the same 
(so $C'$ and $C''$ have the same events, the same conditions and the same flow relation), therefore
we have to check that also the newly added (after/before) inhibitor arcs are the same, i.e.,

\begin{itemize}
\item
$\forall b \in B'$ such that $\post{b} \neq \emptyset$ we have $b \mathrel{\mathcal{A}_1} e \iff b \mathrel{\mathcal{A}_2} e \;$, and 
\item
$\forall b \in \post{e}$ $\forall e' \in E$ we have $\; b \mathrel{\mathcal{B}_1} e' \iff b \mathrel{\mathcal{B}_2} e'$,
\end{itemize}

\noindent
where we denote $\mathcal{A}_1$ (resp. $\mathcal{B}_1$) the after (before) inhibitor arcs
obtained by extending $C$ to $C'$ and $\mathcal{A}_2$ (resp. $\mathcal{B}_2$)
the after (before) inhibitor arcs obtained by extending $C$ to $C''$.
However, these additional requests are trivially satisfied because we know that
$\forall s, s' \in S. (s, s') \in R_1 \Rightarrow (s \in \prei{t_1} \Leftrightarrow s' \in \prei{t_2})$.
In fact, if $b \mathrel{\mathcal{A}_1} e$, then, by \cref{folding-pti--def}, 
there is an inhibitor arc from $\rho_1(b)$ to $t_1$, i.e., $\rho_1(b) \in \prei{t_1}$.
Since $(\rho_1(b), \rho_2(b)) \in R_1$, this implies that $\rho_2(b) \in \prei{t_2}$
and so $b \mathrel{\mathcal{A}_2} e$.
The implication on the other side is symmetrical, and therefore omitted.
The argument for relations $\mathcal{B}_1, \mathcal{B}_2$ is the same, 
and therefore omitted.

To conclude, we have $C' = C''$. 
Thus, $(C, \rho_2) \deriv{e} (C', \rho_2')$ with $\rho_2'(e) = t_2$ and
$(\rho_1'(b),$ $\rho_2'(b)) \in R_1$ for each $b \in \post{e}$. 
Hence, 
for all $b' \in B'$ it holds that $(\rho_1'(b'), \rho_2'(b')) \in R_1$, because
for all $b' \in B$ this holds by hypothesis and for all $b' \in \post{e}$
this follows by construction (thanks to the fact that $(\post{t_1}, \post{t_2}) \in R_1^\oplus$).
As a consequence $\cntp \in R_2$.

The case where $(C, \rho_2)$ moves first is symmetrical and therefore omitted.
Thus, $R_2$ is a causal-net bisimulation and, since $\cntz \in R_2$, we have
$m_1 \sim_{cn} m_2$.
\end{proof}
\end{theorem}

\subsection{It is decidable whether a place relation is a pti-place bisimulation}\label{r_pti_place_bis_decid_sec}
\begin{lemma}\label{r-pti-place-bis-decid-app}
Given a finite PTI net $N = (S, A, T, I )$ and a
place relation $R \subseteq S \times S$,
it is decidable whether $R$ is a pti-place bisimulation.

\begin{proof}
We want to prove that $R$ is a pti-place bisimulation if and only if 
the following two finite conditions are satisfied:
\begin{enumerate}
    \item $\forall t_1$ such that $\pre{t_1} \trns{t_1}$,
          $\forall m$ such that $(\pre{t_1}, m) \in R^\oplus$,
          $\exists t_2$ such that $\pre{t_2} \trns{t_2}$ and
          \begin{alphaenumerate}
          \item $\pre{t_2} = m$,
          \item $(\post{t_1}, \post{t_2}) \in R^\oplus$,
                $l(t_1) = l(t_2)$,
          \item $\forall s, s' \in S. (s, s') \in R \Rightarrow (s \in \prei{t_1} \Leftrightarrow s' \in \prei{t_2})$.
          \end{alphaenumerate}
    \item $\forall t_2$ such that $\pre{t_2} \trns{t_2}$,
          $\forall m$ such that $(m, \pre{t_2}) \in R^\oplus$,
          $\exists t_1$ such that $\pre{t_1} \trns{t_1}$ and
          \begin{alphaenumerate}
          \item $\pre{t_1} = m$,
          \item $(\post{t_1}, \post{t_2}) \in R^\oplus$,
                $l(t_1) = l(t_2)$,
          \item $\forall s, s' \in S. (s, s') \in R \Rightarrow (s \in \prei{t_1} \Leftrightarrow s' \in \prei{t_2})$.
          \end{alphaenumerate}

\end{enumerate}

First we prove the implication from left to right, only for condition 1, as the other is symmetrical.
If $R$ is a pti-place bisimulation and $(\pre{t_1}, m) \in R^\oplus$, then 
from $\pre{t_1} \trns{t_1} \post{t_1}$ it follows that there exists $t_2$ such that  
$\pre{t_2} \trns{t_2} \post{t_2}$ with $\pre{t_2} = m$,
                $(\post{t_1}, \post{t_2}) \in R^\oplus$,
                $l(t_1) = l(t_2)$ and, moreover, $\forall s, s' \in S. (s, s') \in R \Rightarrow (s \in \prei{t_1} \Leftrightarrow s' \in \prei{t_2})$.
Therefore, conditions (a), (b) and (c) are trivially satisfied.
                        
Now we prove the implication from right to left, i.e., if conditions 1 and 2 hold for $R$, then $R$ is a pti-place bisimulation.
Suppose $(m_1, m_2) \in R^\oplus$ and $m_1 \trns{t_1} m_1'$.
Let $ q = \{(s_1, s_1'), (s_2, s_2'), \ldots,$ $(s_k, s_k')\}$ be any multiset of associations
that can be used to prove that $(m_1, m_2) \in R^\oplus$.  So this means that 
$m_1 = s_1 \oplus s_2 \oplus \ldots \oplus s_k$, 
$m_2 = s_1' \oplus s_2' \oplus \ldots \oplus s_k'$
and that $(s_i, s_i') \in R$ for $i = 1, \ldots, k$. 
If $m_1 [t_1 \rangle m_1'$, then $m_1' = m_1 \ominus \pre{t_1} \oplus \post{t_1}$.
Consider the multiset of associations 
$p = \{(\overline{s}_{1}, \overline{s}'_{1})$, 
$\ldots, (\overline{s}_{h}, \overline{s}'_{h})\} \subseteq q$,
with $\overline{s}_{1} \oplus  \ldots \oplus \overline{s}_{h} = \pre{t_1}$.

Note that $(\pre{t_1}, \overline{s}'_{1} \oplus  \ldots \oplus \overline{s}'_{h}) \in R^\oplus$ and that $\pre{t_1} \trns{t_1}$.
Hence, by condition 1, there exists a transition $t_{2}$ such that $\pre{t_2} \trns{t_2}$,
$\pre{t_{2}} = \overline{s}'_{1} \oplus  \ldots \oplus \overline{s}'_{h}$, 
$(\post{t_1}, \post{t_{2}}) \in R^\oplus$, $l(t_1) = l(t_{2})$,  and
$\forall s, s' \in S. (s, s') \in R \Rightarrow (s \in \prei{t_1} \Leftrightarrow s' \in \prei{t_2})$.
By hypothesis, $\prei{t_1} \cap \dom(m_1) = \emptyset$, so
since $(m_1, m_2) \in R^\oplus$ and condition (c) holds, 
we have that $\prei{t_2} \cap \dom(m_2) = \emptyset$.
Therefore, since $\pre{t_2} \subseteq m_2$, 
also $m_2 [t_{2} \rangle m_2'$ is firable, where 
$m_2' = m_2 \ominus \pre{t_{2}} \oplus \post{t_{2}}$, and we have that
$(\pre{t_1}, \pre{t_{2}}) \in R^\oplus$, 
$(\post{t_1}, \post{t_{2}}) \in R^\oplus$, 
$l(t_1) = l(t_{2})$, 
$(m_1 \ominus \pre{t_1}, m_2 \ominus \pre{t_2}) \in R^\oplus$ and, moreover, $\forall s, s' \in S. (s, s') \in R \Rightarrow (s \in \prei{t_1} \Leftrightarrow s' \in \prei{t_2})$,
as required, where $(m_1 \ominus \pre{t_1}, m_2 \ominus \pre{t_2}) \in R^\oplus$ holds as, from the set $q$
of matching pairs for $m_1$ and $m_2$,
we have removed those in $p$.

If $m_2 [t_2 \rangle m_2'$, then we have to use an argument symmetric to the above, where condition 2 is used instead.
Hence, we have proved that conditions 1 and 2 are enough to prove that $R$ is a pti-place bisimulation.

Finally, the complexity of this procedure is as follows. For condition 1, 
we have to consider all the net transitions,
and for each $t_1$ we have to consider
all the markings $m$ such that $(\pre{t_1}, m) \in R_i^\oplus$, and for each pair $(t_1, m)$
we have to check whether there exists a transition $t_2$ such that $m = \pre{t_2}$, $l(t_1) = l(t_2)$, 
$(\post{t_1}, \post{t_2}) \in R_i^\oplus$ and, moreover, that $\forall s, s' \in S. (s, s') \in R \Rightarrow (s \in \prei{t_1} \Leftrightarrow s' \in \prei{t_2})$. And the same for condition 2. Hence, this  
procedure has worst-case time complexity
$O(q \cdot n^{p} \cdot q \cdot  (p^2\sqrt{p} + n^2 \cdot p))$,
where $q = |T|$, $n = |S|$ and  $p$ is the least number such that 
$|\pre{t}| \leq p$,  $| \prei{t}| \leq p$,
and $|\post{t}| \leq p$ for all $t \in T$, 
as the number of markings $m$ related via $R_i$ to $\pre{t_1}$ is  $n^{p}$ at most, 
checking whether $(\post{t_1}, \post{t_2}) \in R_i^\oplus$ takes
$O(p^2\sqrt{p})$ in the worst case
and, finally, checking the conditions on the inhibiting sets
is $n^2 \cdot p$ at most.
\end{proof}
\end{lemma}

\end{document}